\documentclass[%
 reprint,
 amsmath,amssymb,
 aps,
 showkeys,
prxlife,
]{revtex4-2}

\usepackage{graphicx}
\usepackage{dcolumn}
\usepackage{bm}
\usepackage{xcolor}
\usepackage[T1]{fontenc}
\usepackage[normalem]{ulem}
\usepackage{amssymb}
\usepackage{amsmath}
\usepackage{ulem}
\usepackage{cancel}

\newcommand{\numin}[0]{\nu^*_\mathrm{min}}
\newcommand{\numax}[0]{\nu^*_\mathrm{max}}
\newcommand{\nuread}[0]{\nu^*_\mathrm{read}}
\newcommand{\Fmin}[0]{F^*_\mathrm{min}}
\newcommand{\Fmax}[0]{F^*_\mathrm{max}}
\newcommand{\dthetaRP}[0]{\Delta\theta_\mathrm{RP}}
\newcommand{\dRRP}[0]{\Delta R_\mathrm{RP}}
\newcommand{\btheta}[0]{{\bm{\theta}}}
\newcommand{\bR}[0]{{\bm{R}}}
\newcommand{\bnabla}[0]{{\bm{\nabla}}}

\newcommand{\norm}[1]{\left\lVert#1\right\rVert}
\newcommand{\rev}[2]{\textcolor{black}{#2}}

\begin{document}


\title{Learning by training: emergent physical memory from cyclically tuning disordered sphere packings}

\author{Mengjie Zu}
\author{Carl P. Goodrich}%
\email{carl.goodrich@ist.ac.at}
\affiliation{%
 Institute of Science and Technology Austria (ISTA), Am Campus 1, 3400 Klosterneuburg, Austria
}%

\date{\today}

\begin{abstract}
Many living and artificial systems improve their fitness or performance by adapting to changing environments or diverse training data. However, it remains unclear how\rev{ such}{} environmental variation \rev{influences}{shapes} adaptation, what is learned\rev{ in the process}{}, and \rev{whether}{when} memory of past conditions is retained. 
\rev{In this work, we investigate these questions using athermal disordered systems that are subject to cyclic inverse design, enabling them to attain target elastic properties spanning a chosen range. We demonstrate that such systems evolve toward a marginally absorbing manifold (MAM), which encodes memory of the training range that closely resembles return-point memory observed in cyclically driven systems. We further propose a general mechanism for the formation of MAMs and the corresponding memory that is based on gradient discontinuities in the trained quantities. Our model provides a simple and broadly applicable physical framework for understanding how adaptive systems learn under environmental change and how they retain memory of past experiences.}{Here, we show how cyclic environmental change can produce robust memory. Using a model athermal disordered solid trained by inverse design to attain target elastic properties over a prescribed range, we find that the system evolves toward a marginally absorbing manifold (MAM), meaning that training is reversible within the training range but not beyond it, which encodes a memory of that range.
We further propose a general mechanism for MAM formation and memory encoding based on discontinuities in the gradient of the trained quantity. These results provide a simple, broadly applicable physical framework for how adaptive systems learn under changing environments and retain memory of past conditions.}


\end{abstract}


\maketitle

\section{Introduction}

From evolution to artificial neural networks to nearly any inverse-design problem, there is a staggering list of systems where internal variables are iteratively adjusted to optimize one or more emergent features.
The process of optimization can vary dramatically in these systems, for example machine learning is predicated on global gradients~\cite{duda1973pattern,ackley1985learning,rumelhart1985learning,lecun2002gradient,krizhevsky2012imagenet,lecun2015deep,ruder2016overview,bottou2018optimization} while many physical systems approximate gradient descent using only local information~\cite{scellier2017equilibrium,pashine2019directed,stern2021supervised,stern2023learning,lopez2023self,anisetti2023learning,anisetti2024frequency,dillavou2022demonstration,stern2022physical,altman2024experimental,dillavou2024machine,tang2024learning,falk2025temporal,arinze2023learning,evans2024pattern,PhysRevResearch.7.013157, behera2023enhanced, veenstra2025adaptive,dillavou2025understanding}. 
Many systems, however, also include fixed or slowly varying ``environmental variables'' that do not evolve through the same optimization process. These could include the presence of a predator that affects an organism’s optimal running speed, the data used to train an LLM, or a desired self-assembly outcome. Such quantities clearly play a critical role in optimization, and if they change, so too will the optimal solution.

\rev{However, we lack a broad and generalized understanding of the effects of environmental change. For example, when can a system retain memory of its past environment? This is at least plausible for under-constrained systems where optimal solutions are degenerate, since this degeneracy can be exploited to reflect traces of a system’s history. Can this actually happen in practice? How does it work? What is remembered and how is the memory stored/retrieved? In this paper, we show that cyclic environmental changes can lead to precise memory formation, and uncover a general mechanism for how this works. }
{However, we still lack a broad understanding of how environmental variation affects optimization over time. In particular, when can a system retain memory of past environments, what is remembered, and how is that memory stored and read out? 
To address these questions, we show that cyclic environmental variation during training can drive a system to a so-called marginally absorbing manifold (MAM) that stores a memory of the training history, and we uncover a broadly applicable mechanism for how this occurs.}

We study \rev{such}{} memory formation in the context of athermal disordered sphere packings, which we optimize, or ``train,'' to control their elastic constants~\cite{zu2024designing-57d,zu2025fully-43b}. We systematically alter the environment by cyclically changing the target elastic constant between two chosen values\rev{}{, which we refer to as the training range.}
We find that the system often falls into an absorbing manifold, returning to the same state after every cycle~\cite{hirsch1977invariant}\rev{. However, this manifold is only marginally absorbing, meaning that the system does not return to the same state if trained outside of the initial training range. The fact that the system finds a marginally absorbing manifold (MAM) allows one to ``read out'' both ends of the training range, meaning that the system has stored this information in memory.}{, but that this manifold is only marginally absorbing: training remains reversible within the original range, but not if the target is driven beyond it. This allows both ends of the training range to be read out, indicating that the system has stored them in memory.}
We then present a general mechanism for the formation of MAMs based on discontinuities in the gradient of the quantity being trained.
This \rev{theory}{mechanism} is not specific to our \rev{system and has the potential to explain nontrivial learning caused by training in a changing environment in a wide range of settings.}{sphere packings and provides a route to nontrivial memory formation under changing environments in a broad class of systems.}

\rev{
The memory we observe is a return-point memory (RPM), a form of memory in which a system recalls previously visited states and returns to them under the right external drive. 
RPM has been used to describe magnetic hysteresis~\cite{barker1983magnetic-5b8,sethna1993hysteresis-05e,preisach1935ber-c74}, and is observed in a variety of cyclically sheared systems~\cite{RevModPhys.91.035002,paulsen2025mechanical-d8c,paulsen2019minimal-36e,PhysRevLett.113.068301,PhysRevLett.112.025702,keim2013multiple-1d6,adhikari2018memory-84e,mungan2019networks-3f5,lindeman2021multiple,scheff2021actin-98a}. For example, 
}{
The memory we achieve is related to that observed in a wide variety of other physical systems that encode aspects of their past history~\cite{RevModPhys.91.035002, paulsen2025mechanical-d8c, mungan2025self-organization-787}. Some of the many such systems include 
charge density waves~\cite{gill1981transient-418,fleming1983transient-3e7}, 
crumpled sheets~\cite{matan2001crumpling-c0f, PhysRevLett.118.085501},
sheared glasses and jammed solids~\cite{bertin2003kovacs-554, cugliandolo2004memory-91e,PhysRevLett.112.025702, PhysRevLett.112.028302,mungan2019networks-3f5,PhysRevResearch.2.012004, adhikari2018memory-84e,lindeman2021multiple,regev2021topology-1b4,PhysRevE.96.020101,PhysRevE.58.4673, PhysRevLett.122.158001, royer2015precisely,PhysRevLett.110.018302,PhysRevLett.93.088001,fiocco2015memory,chen2023memory-e21,chen2025microstructural-ab0}, suspensions~\cite{pine2005chaos-3f0,Corté2007,menon2009universality-967,PhysRevLett.107.010603,keim2013multiple-1d6,pham2015particle-ebf,PhysRevLett.113.068301}, and actin networks~\cite{scheff2021actin-98a}. 
The phenomenology of cyclically sheared suspensions of particles provides a particularly relevant example for us: after repeated shearing, the system
}\rev{a cyclically sheared suspension of particles}{}
can reach an absorbing state where particles no longer collide, making the trajectories permanently reversible\rev{~\cite{pine2005chaos-3f0,Corté2007,menon2009universality-967}}{}. However, if a system is subsequently driven beyond the initial maximum strain amplitude\rev{ $\gamma_0$}{}, particle collisions occur.
In this way, the material encodes a memory of the initial strain amplitude. 
Our results show that such behavior is possible in systems governed by optimization dynamics, where memory is learned by cyclic training rather than cyclic shear\rev{}{, and provide an explanatory mechanism for such behavior}.

The paper is organized as follows. In Section~\ref{sec:model}, we present our model for trainable disordered sphere packings and discuss our protocol for cyclic training. Section~\ref{sec:numerical_results} details our observation that cyclically trained sphere packings approach a marginally absorbing manifold with\rev{ return-point}{} memory, and discusses the nontrivial role of contact changes during training. Section~\ref{sec:mechanism} identifies the crucial feature of a contact change as presenting a discontinuity in the gradient of the property being trained, leading to sharp changes in the training dynamics, and develops a theory called Gradient Discontinuity Learning (GDL) that explains how any trainable system with gradient discontinuities can, in principle, reach a MAM. Finally, we discuss these results in Section~\ref{sec:discussion}, and speculate on the applicability of GDL to a variety of different systems across fields.

\section{Cyclic training in disordered sphere packings \label{sec:model}}
We consider two-dimensional athermal packings of $N=128$ particles with equal mass $m$. The particles are evenly divided into $n_{sp}=32$ species, with all particles of species $\alpha$ having diameter $D_\alpha$. 
Particles $i$ and $j$ interact via a purely repulsive, short-ranged Hertzian potential, given by
\begin{align} \label{eq:hmorse}
    U_{ij}(r_{ij}) = \begin{cases}
        \frac{k}{2.5}(1 - r_{ij}/\sigma_{ij})^{2.5}, & r_{ij} < \sigma_{ij} \\
        0, & r_{ij} \geq \sigma_{ij},
    \end{cases}
\end{align}
where $r_{ij}$ is the center-center distance between the particles, $\sigma_{ij}$ is the mean of their diameters, and $k$ determines the energy scale. We set $m=k=1$. 

Initial systems are prepared by setting half the $D_\alpha$ to $1$ and half to $1.4$, creating a 50/50 mixture of small and large particles to avoid crystallization. 
Particles are then randomly placed within a square periodic simulation box of length $L$, which is set to achieve an area fraction of $\phi_0= 0.95$. We use the FIRE algorithm~\cite{bitzek_structural_2006,guenole_assessment_2020} to obtain a mechanically stable athermal solid by quenching the potential energy to a local energy minimum.

\begin{figure*}
    \centering
    \includegraphics[width=0.8\linewidth]{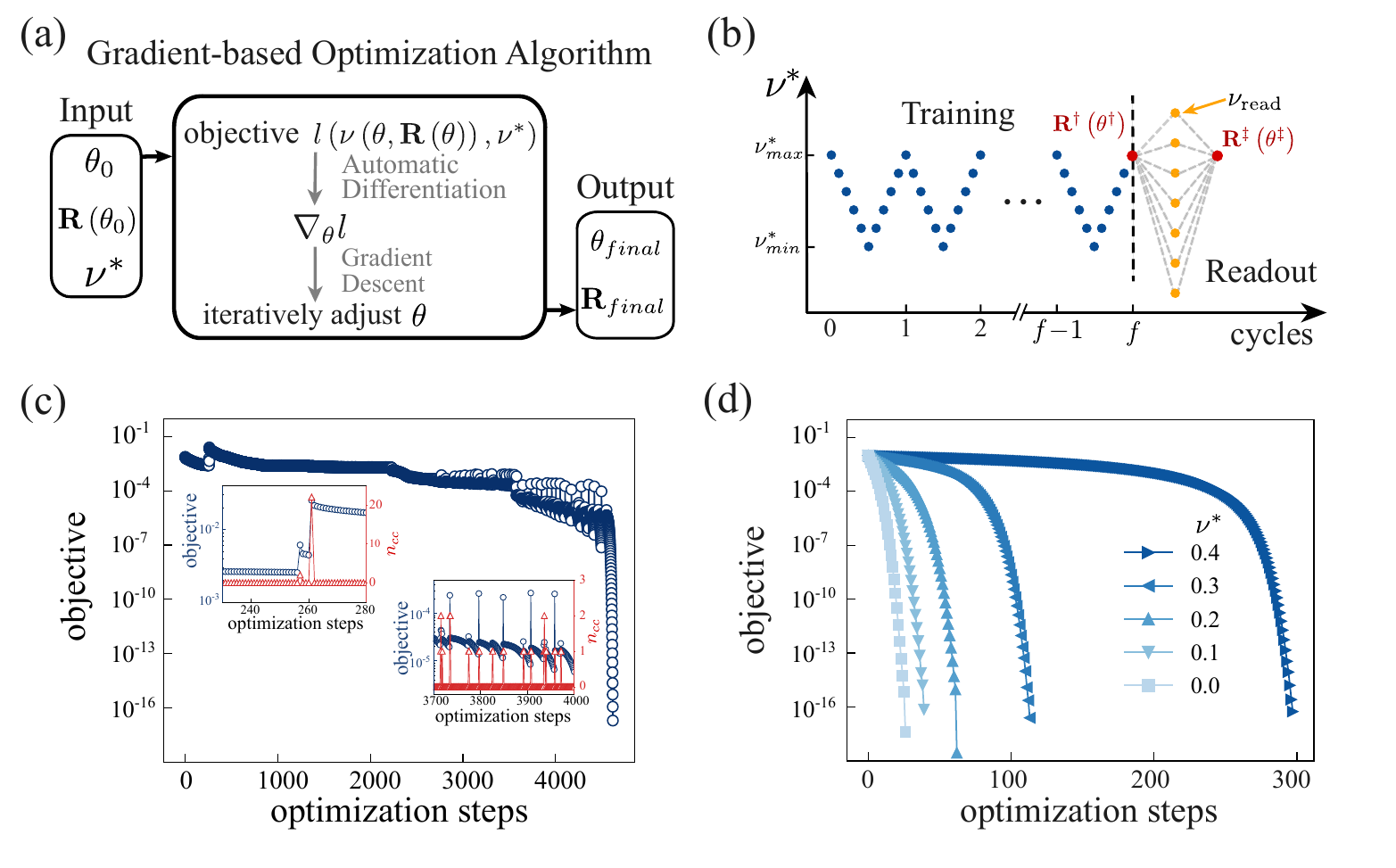}
    \caption{Training and cyclic training.
    (a) A schematic of our gradient-based optimization routine to tune the Poisson's ratio $\nu$ toward a target value $\nu^*$, which includes input, output, and a gradient descent algorithm that iteratively modifies the parameters $\btheta$. Other elastic constants are trained in the same way.
    (b) A sketch of cyclic training. We start by training to $\numax$ before cyclically adjusting the target $\nu^*$ between $\numax$ and $\numin$. 
    Each blue point refers to an individual training as described in (a). After $f$ complete cycles (first red point), the positions $\bR^\dagger$ and parameters $\btheta^\dagger$ form the initial inputs for readout measurements. A set of parallel readout trainings are conducted first with a wide range of $\nu^*=\nuread$ (orange points), and then with $\nu^*=\numax$, generating the final positions $\bR^\ddagger$ and parameters $\btheta^\ddagger$ (second red point). 
    (c) The objective $\ell$ for a representative example of the initial first training to $\nu^*= 0.5$. The optimization is successful, converging in just over 4500 steps. The insets zoom in on regions where $\ell$ exhibits sharp changes and highlight their correspondence with contact changes. 
    (d) The objective $\ell$ during a half-cycle of training following 23 complete training cycles in which the system approaches a MAM. The five curves correspond to sequential training steps: from $\nu=0.5$ to $\nu^*=0.4$, then from $\nu=0.4$ to $\nu^*=0.3$, and so on, down to $\nu^*=0.0$. }
    \label{fig:methods}
\end{figure*}

Following the methods presented in Refs.~\cite{zu2024designing-57d,zu2025fully-43b}, we then train various mechanical properties of the system by iteratively adjusting a chosen set of parameters, $\btheta$. In this work, we set $\btheta$ to be the $n_\mathrm{sp}$ species-level particle diameters, $\left\{D_\alpha\right\}$, but we refer to them as $\btheta$ to emphasize that we \textit{could} consider only a subset of them, or include other variables such as energy scales. 
We focus primarily on training the Poisson's ratio $\nu$, but also consider other elements of the elastic modulus tensor \rev{}{(see Appendix~\ref{sec:cijkl_calculation})}. 

Our training routine is sketched in Fig.~\ref{fig:methods}a. We start with an initial set of parameters, $\btheta_0$, a set of particle positions corresponding to a local energy minimum, $\bR(\btheta_0)$, and a desired target Poisson's ratio, $\nu^*$. 
The training routine calculates the current value of the Poisson's ratio, $\nu(\btheta,\bR(\btheta))$, and compares it to $\nu^*$ to construct an objective function $\ell(\btheta) = (\nu(\btheta, \bR(\btheta))-\nu^*)^2$. Here, we have made the $\btheta$-dependence explicit. We then employ methods of Automatic Differentiation~\cite{baydin2018automatic,10.1038/323533a0,10.1145/355586.364791} to calculate the gradient $\bnabla_\btheta \ell$, which is fed into a standard gradient descent routine that iteratively adjusts $\btheta$ (and $\bR(\btheta)$), until either $\ell$ drops below $10^{-16}$ or a maximum number of iterations has been reached. During each iteration, the energy of the system is re-minimized with the current $\btheta$. More details can be found in Appendix~\ref{sec:optimization_algorithm}.
The output of the routine is the final set of parameters and particle positions. 

This training routine was sufficiently developed in Refs.~\cite{zu2024designing-57d,zu2025fully-43b} that, for our purposes, we can consider it a black box for the inverse design of disordered solids. Here, we are interested in exploring the \textit{effects} that such training has on the system beyond simply controlling the target quantity (\textit{e.g.} the Poisson's ratio). To explore this, we \textit{cyclically} train a system, following a protocol outlined in Fig.~\ref{fig:methods}b. Using the Poisson's ratio $\nu$ as our primary example, we iteratively and incrementally train the system between $\numax$ and $\numin$, which are chosen in advance. In the first example in the next section, for instance, we first train at $\nu^*=\numax=0.5$. Then, we use the final $\btheta$ and $\bR(\btheta)$ as inputs and train with $\nu^*=0.4$, continuing in this manner with $\nu^*=0.3$, etc., until $\nu^*=\numin=0.0$. We then reverse course, setting $\nu^*=0.1$, and then $\nu^*=0.2$, until we return to $\nu^*=\numax$. This constitutes a single cycle, and we repeat this process for $f$ cycles. 

\rev{}{This protocol is related in spirit to the oscillatory training strategy of Falk et al.~\cite{falk2023learning}, who showed that switching between incompatible targets selects for adaptable regions of design space. We find that this produces not only adaptable solutions but a marginally absorbing manifold that encodes memory of the training range.}
\rev{W}{To observe this, w}e also occasionally perform an array of parallel ``readout'' calculations. After some number of complete cycles, with $\nu^*=\numax$, we first train with $\nu^*=\nu^*_\mathrm{read}$, and then train back to $\nu^*=\numax$. This is repeated, always starting with the same initial $\btheta$ and $\bR(\btheta)$, for a range of $\nu^*_\mathrm{read}$ that is not limited to being between $\numin$ and $\numax$. We will describe the precise measurements and insights that we obtain from these readout calculations below. 

Finally, note that this cyclic training protocol is computationally intensive. We typically train for 10-30 cycles, with each training cycle consisting of around 10 distinct optimizations. Each optimization typically requires up to $\sim 10^{4}$  optimization steps to converge, with each step requiring a highly accurate energy minimization, which itself needs up to $\sim 10^{3}$ steps of the FIRE algorithm, and as many force calculations, to converge. To partially mitigate this, during cyclic training we use a variable learning rate and set the maximum number of optimization steps to be $10^{3}$ meaning that we do not always reach the target $\nu^*$ \rev{}{during the first few training cycles}. We also consider relatively small systems, with $N=128$ particles, although Refs.~\cite{zu2024designing-57d,zu2025fully-43b} show that the training of $\nu$ is unaffected by system size.

\section{Cyclic training encodes a memory
\label{sec:numerical_results}}
\subsection{A first observation}
Figure~\ref{fig:methods}c shows the objective function, $\ell$, for the first iteration of training from a random initial state with $\nu=0.587$ to one with $\nu = \nu^* = 0.5$. This typical training trajectory converges after approximately 4500 optimization steps, and highlights two behaviors that are important to understand. First, around the 260th optimization step, the system undergoes a structural rearrangement where the minimum $\bR(\btheta)$ becomes a saddle point, resulting in a large-scale change in the system that causes $\ell$ to increase significantly and about 20 contacts to change (see first inset). Second, throughout the latter half of the optimization, $\ell$ exhibits occasional, almost periodic, spikes~\footnote{Note that these spikes are not observed in Ref.~\cite{zu2024designing-57d}. As will become clear in Section~\ref{sec:mechanism}, this is due to the smooth attractive potential used in that paper.} (see second inset). The insets also show the number of particle-particle contact changes at each step.
The spikes in $\ell$  occur when 1 or 2 contacts change, but are not associated with large-scale structural rearrangements. 

This is in stark contrast to what we observe after numerous training cycles. For example, the dark blue curve in Fig.~\ref{fig:methods}d shows the first training trajectory of the 24th cycle, where $\nu$ is trained from $0.5$ to a target of $\nu^*=0.4$. Subsequent curves show the training trajectories for the rest of this half cycle, ending with $\nu^*=0$. These trajectories are qualitatively different. First, $\ell$ decreases smoothly and continuously: there are no spikes or sharp features of any kind, nor are there any structural rearrangements or even contact changes! This is already surprising, as the existence of a configuration that can span a wide range in $\nu$ without changing the contact topology is not trivial. Second, training is significantly ``easier,'' in that the number of required optimization steps decreases by 1-2 orders of magnitude. 

Clearly, something dramatic beyond the Poisson's ratio has changed during cyclic training. 
We will see that after a sufficient number of training cycles, the trajectory of the parameters reaches an absorbing manifold, where the same configurations and parameters are revisited every cycle. 
\rev{The word manifold refers to the path in $\btheta$-space during a cycle of training, and an absorbing manifold is one that returns to the same place: once the system finds an absorbing manifold, it cannot leave it. }
{An absorbing manifold refers to any region in $\btheta$-space that spans a certain range in $\nu$ (the quantity being trained), such that the parameters remain on this manifold when training within this range.}
Furthermore, this manifold is marginally absorbing, meaning that it becomes non-absorbing \rev{}{(the parameters do not return to the manifold)} if trained beyond the range defined by $\numin$ and $\numax$. Thus, such a marginally absorbing manifold (MAM) encodes into memory the values of $\numin$ and $\numax$. These memories can be read out as jumps or kinks in various quantities, including 1) return-point changes in parameters, 2) return-point changes in particle positions, 3) the number of required optimization steps, and 4) 
\rev{a particular component of the change in parameters after training}{a particular measure of the change in parameters, defined in Eq.~\eqref{eq:theta_perp} below.} 
The rest of this section will detail and quantify these observations, and discuss the role of contact changes. In Section~\ref{sec:mechanism}, we will build a theory based on the effect of contact changes that can generically predict and explain how a system can learn a specific MAM through cyclic training.

\begin{figure*}
    \centering \includegraphics[width=1.0\linewidth]{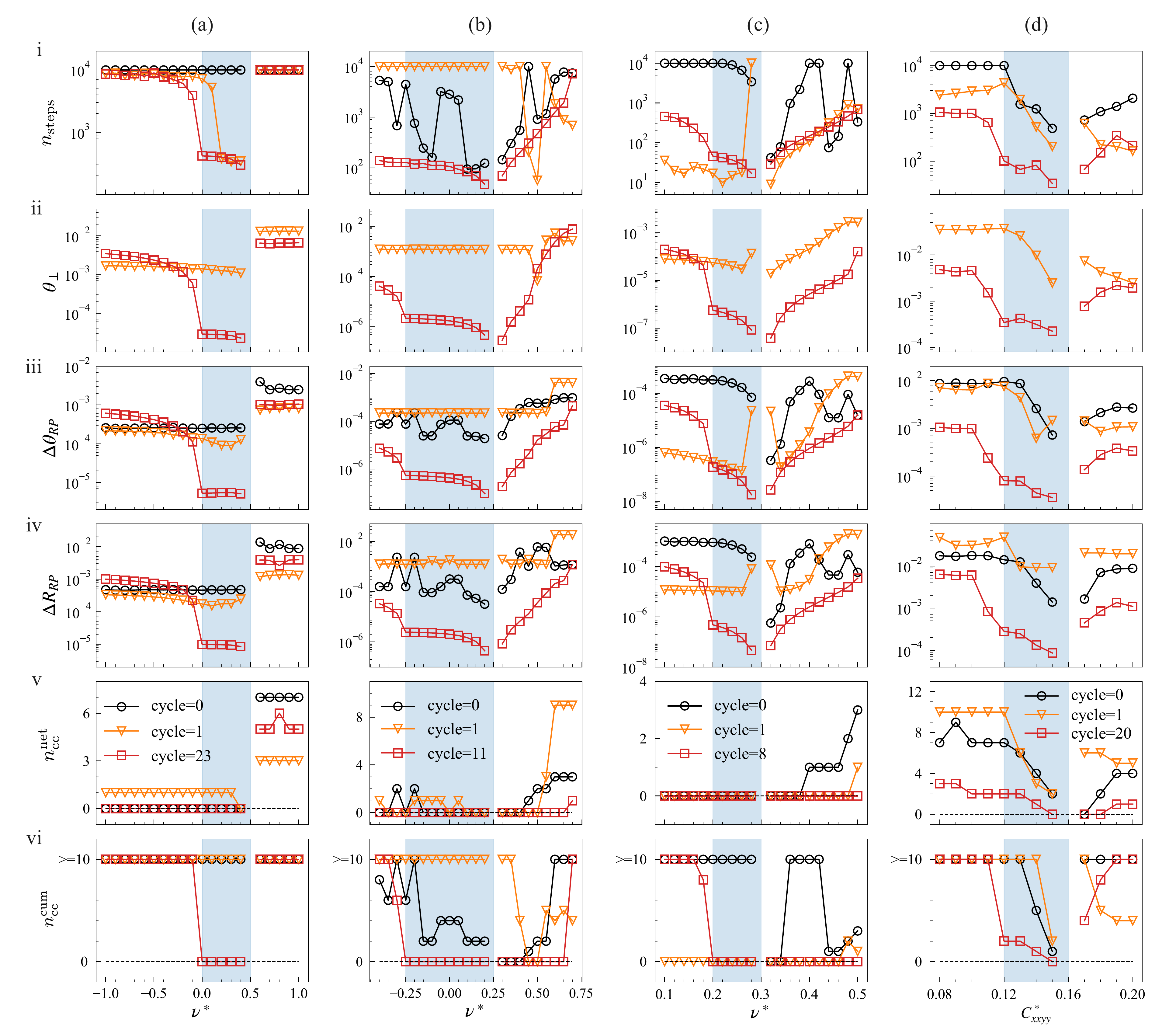}
    \caption{Memory in four representative examples. Columns (a)–(c) show readout results following cyclic training targeting the Poisson’s ratio over different target ranges, while column (d) corresponds to targeting the (arbitrarily chosen) elastic constant $c_{xxyy}$. The target training ranges are indicated by gray shading. The initial values of the mechanical properties are $\nu_0=0.587$, $\nu_0=0.681$, $\nu_0=0.714$, and $c_{xxyy,0}=0.157$, respectively. All readout is obtained following the procedure described in the text, and begins from a trained state at $\numax=0.5$, $\numax=0.25$, $\numax=0.3$, and $c^*_{\mathrm{xxyy,max}}=0.16$, respectively.
    Each row displays a different readout measurement: (i) the number of optimization steps, $n_{\text{steps}}$; (ii) the perpendicular distance $\theta_{\perp}$ to the line defined by $\boldsymbol{\vartheta}$; (iii) return-point changes in the parameters, $\dthetaRP$; (iv) return-point changes in particle positions, $\dRRP$; (v) the net number of contact changes  $n^{\text{net}}_{\text{cc}}$; and (vi) the cumulative number of contact changes $n^{\text{cum}}_{\text{cc}}$ measured during readout training with a given $\nu^*_\mathrm{read}$.
    Data is shown for readout measurements performed after 0 training cycles (but after training to $\numax$ or $c^*_{\mathrm{xxyy,max}}$) (black), after 1 complete training cycle (orange), and after the system reaches a MAM (red). 
     }
    \label{fig:memory}
\end{figure*}

\begin{figure}
    \centering
    \includegraphics[width=\linewidth]{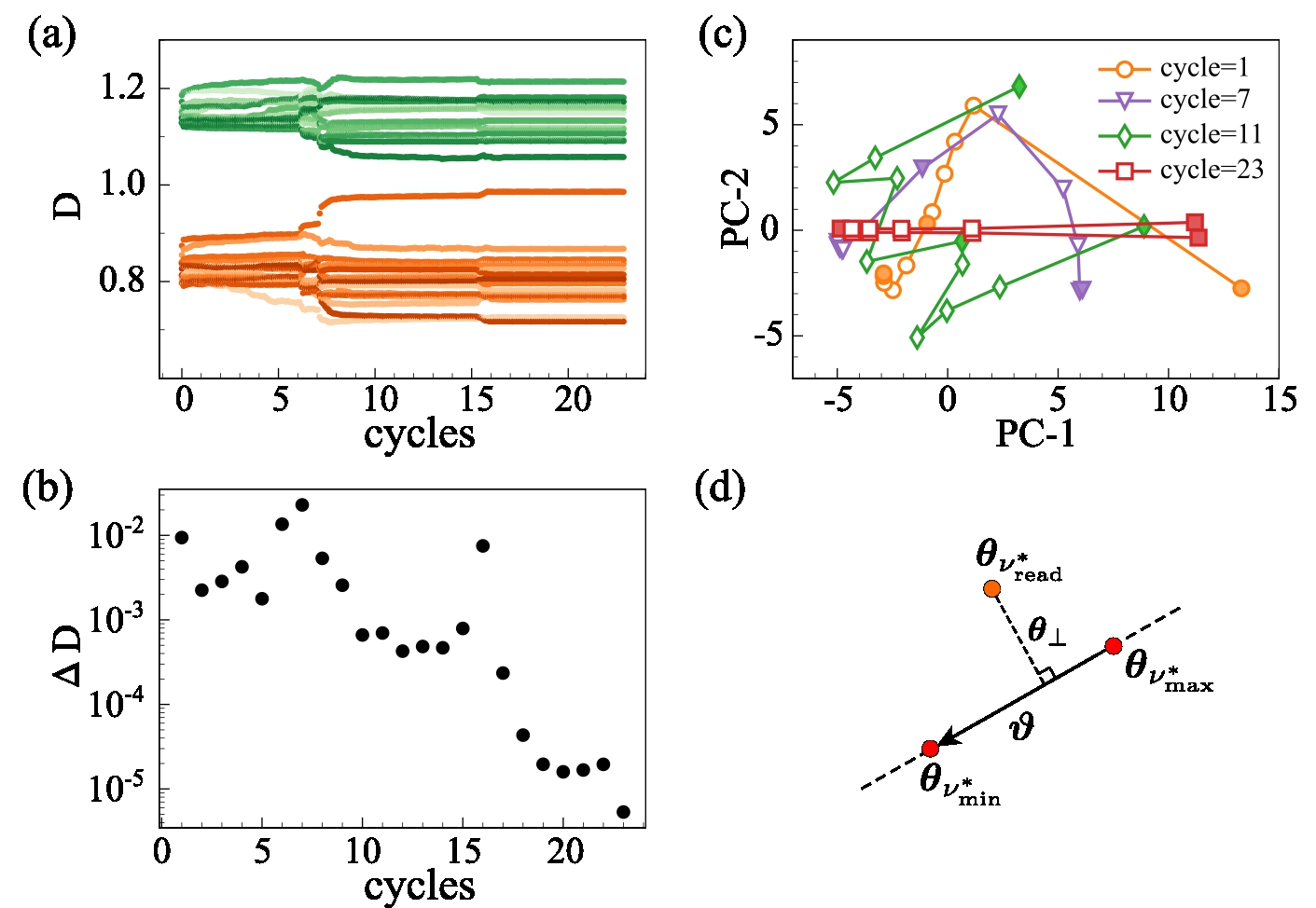}
    \caption{
    The evolution of the parameters during cyclic training. 
    (a) The 32 species-level diameters over 23 training cycles. 
    (b) The magnitude of the change in parameters from the beginning to the end of each cycle, $\Delta D$, decreases as the system approaches a MAM.
    (c) Projection of the parameter vectors onto the first two \rev{principle}{principal} components, shown for a few training cycles. Data points from the same completed cycle are connected by lines, with the starting and ending points marked by solid symbols. After 23 cycles, the parameters appear close to collinear, aligning with the primary \rev{principle}{principal} component. 
    (d) Sketch of $\bm{\vartheta}$ and $\theta_\perp$. We define the vector $\bm{\vartheta}$ to point from the parameters at $\numax$ to those at $\numin$ from the last half training cycle. $\theta_\perp$ is the distance to the line defined by this vector. 
    }
    \label{fig:parameter}
\end{figure}

\subsection{Emergent memory after cyclic training}

We now discuss a number of quantitative readout measurements, shown in Fig.~\ref{fig:memory}, that demonstrate the MAM as well as the emergence of memory.
Readout measurements are performed before the first training cycle but after the initial optimization to $\numax$ (black), after the first complete training cycle (orange), and after the system converges to a MAM (red). The shaded gray regions indicate the training range defined by $\numin$ to $\numax$. We will begin with the left column, which shows a representative system with the Poisson's ratio trained between $\numin=0$ and $\numax=0.5$. The other columns, which show different target ranges and mechanical properties, will be discussed later. 

\subsubsection{Ease of training}
We begin by quantifying the apparent speedup in training observed in Fig.~\ref{fig:methods}c-d. Figure~\ref{fig:memory}a(i) shows the number of optimization steps, $n_\mathrm{steps}$, to train the system from $\numax$ to $\nuread$. 
Note that since we always start at $\numax$, readout data for $\nuread=\numax$ is meaningless and are not shown. Importantly, for readout calculations, we train with a fixed learning rate of $10^{-6}$, while in normal cyclic training we use a variable learning rate (see Appendix~\ref{sec:cyclic_training_appendix}). This enables a more quantitative comparison of $n_\mathrm{steps}$ for different $\nuread$. We terminate training after $10^4$ steps, so this should be interpreted as a bound of the true number of steps. 

Figure~\ref{fig:memory}a(i) shows that before cyclic training, $n_\mathrm{steps}=10^4$, meaning that the optimization has not yet converged. After a single training cycle, $n_\mathrm{steps}$ drops below $10^3$ for some of the $\nuread$, while after 23 cycles, $n_\mathrm{steps}$ is below $10^3$ for all the values of $\nuread$ between $\numin$ and $\numax$. However, even after 23 cycles, $n_\mathrm{steps}$ jumps by over a decade as soon as $\nuread$ leaves this range. Apparently, whatever makes training ``easier'' after many cycles does not strongly affect training outside this range.

\subsubsection{Training dynamics}
To better understand what makes training easier after 23 cycles, Fig.~\ref{fig:parameter}(a) shows how the parameters (which again are species-level particle diameters) evolve during the 23 training cycles. While there are occasionally clear and dramatic parameter changes, \textit{e.g.} during the 7th and 8th cycles, they eventually reach a steady state and change very little.
Figure~\ref{fig:parameter}(b) shows the change in the parameters between successive cycles, $\Delta D_k=\sqrt{\frac{1}{n_\mathrm{sp}}\sum_{\alpha=1}^{n_{sp}}\left(D_{\alpha,k}-D_{\alpha,k-1}\right)^2}$, where $k$ indicates the cycle and the sum is over the $n_\mathrm{sp}$ species. This decreases by three orders of magnitude over time, indicating convergence of the cyclic training process and suggesting that the system is progressively approaching an absorbing manifold.
To better visualize the parameter changes, we perform a \rev{Principle}{Principal} Component Analysis on the data in Fig.~\ref{fig:parameter}a. Figure~\ref{fig:parameter}c shows the parameters projected onto the largest two \rev{principle}{principal} components for a few select cycles. Unsurprisingly, only for the 23rd cycle, after the system reaches the absorbing manifold, do the parameters return very nearly to where they began. 

What is perhaps surprising, however, is that the parameters during the 23rd cycle appear to be nearly co-linear. Apparently, once on the absorbing manifold, the training dynamics only explores a single degree of freedom when tuning continuously between $\numin$ and $\numax$.
To show that this is not a consequence of the particular projection in Fig.~\ref{fig:parameter}c, we define $\theta_\perp$ as the distance (in parameter space) from a given set of parameters $\boldsymbol \theta$ to the line that connects the parameters at $\numax$ and $\numin$ from the last half cycle of training, see Fig.~\ref{fig:parameter}d. More precisely, given these reference parameters $\boldsymbol{\theta}_{\numax}$ and $\boldsymbol{\theta}_{\numin}$, then 
\begin{equation}
\theta_{\perp}=
\norm{\left(\boldsymbol{\theta}-\boldsymbol{\theta}_{\numax}\right)-
\frac{\left(\boldsymbol{\theta}-\boldsymbol{\theta}_{\numax}\right)\cdot \boldsymbol{\vartheta}}{\Vert \boldsymbol{\vartheta}\Vert^2}\cdot \boldsymbol{\vartheta},
} \label{eq:theta_perp}
\end{equation}
where
\begin{equation} \label{eq:vartheta}
    \boldsymbol{\vartheta} = \boldsymbol{\theta}_{\numin} - \boldsymbol{\theta}_{\numax}.
\end{equation}

Figure~\ref{fig:memory}a(ii) shows $\theta_\perp$ obtained during the readout calculations. Within the training range, $\theta_\perp$ is less than $10^{-4}$, indicating that indeed these parameters are very nearly colinear. \rev{}{This is likely connected to the results of Falk et al.~\cite{falk2023learning}, who show significantly increased adaptability after cyclicly training elastic networks.} However, this \rev{}{colinearity} breaks down immediately outside the training range. 
While the data is not perfectly correlated, this at least informs the apparent ease of training within the absorbing manifold. 



\subsubsection{\rev{Return-point memory}{Reversible training}}
$n_\mathrm{steps}$ and $\theta_\perp$ both show clear features at $\nuread=\numin$ and $\nuread=\numax$, indicating memory. We now define two ``return-point'' measurements \rev{}{that compare the system at $\numax$ before and after intermediate training to $\nuread$}, both of which also exhibit memory. 
First, we define
\begin{equation}
    \dthetaRP = \sqrt{\frac 1{n_\mathrm{sp}}\sum^{n_\mathrm{sp}}_{\alpha=1}\left(\theta^{\dagger\dagger}_\alpha - \theta^\dagger_\alpha\right)^2},
\end{equation}
where the sum runs over the $n_\mathrm{sp}$ parameters, and $\theta^{\dagger}$ and $\theta^{\dagger\dagger}$ are the parameters before and after readout training, respectively (Fig.~\ref{fig:methods}b). 
$\dthetaRP$ quantifies how the parameters change from before readout training, where $\nu = \numax$, to after readout training, where $\nu$ again equals $\numax$, so that if the training was completely reversible, then $\dthetaRP=0$. As shown in Fig.~\ref{fig:memory}a(iii), $\dthetaRP$ is never strictly $0$ but is quite small ($\dthetaRP<10^{-5}$) after cyclic training for $\numin \leq \nuread \leq \numax$. \rev{}{Note that the nonzero parameter changes are not caused by imperfect convergence during training, but rather 
are influenced by the learning rate, see Appendix~\ref{sec:analysis_nonzero_changes}.} However, even after 23 cycles there is a dramatic increase in $\dthetaRP$ outside of this range. 

Similarly, we define 
\begin{equation}
    \dRRP = \sqrt{\frac 1N \sum_{i=1}^{N}\left(R^{\dagger\dagger}_i - R^\dagger_i\right)^2},
\end{equation}
which quantifies the same return-point change but in the particle positions rather than in the parameters. Figure~\ref{fig:memory}a(iv) shows that the behavior of $\dRRP$ is qualitatively identical to that of $\dthetaRP$.
Together, these results show that the readout training is almost completely reversible for $\numin \leq \nuread \leq \numax$, but not reversible otherwise. The sharp jumps in these quantities clearly indicate two-sided \rev{return-point }{}memory, and are very reminiscent of the \rev{return-point }{}memory observed in a variety of cyclically sheared systems~\cite{RevModPhys.91.035002,paulsen2025mechanical-d8c,mungan2025self-organization-787,pine2005chaos-3f0,Corté2007,menon2009universality-967,PhysRevLett.107.010603,keim2013multiple-1d6,pham2015particle-ebf,PhysRevLett.113.068301}.

\subsubsection{MAMs with different training ranges and target quantities}
Figure~\ref{fig:memory}a(i-iv) demonstrates that cyclic training leads to a MAM, thus storing memory of $\numin$ and $\numax$. Importantly, this result is not specific to this one example system or the choice of $\numin$ and $\numax$. Figure~\ref{fig:memory}b and c show similar data for systems with different training ranges for $\nu$, first with a lower range ($\numin=-0.25$, $\numax=0.25$), and then with a smaller range ($\numin=0.2$, $\numax=0.3$). Finally, Fig.~\ref{fig:memory}d shows the results for cyclically training a different elastic constant, specifically $c_{xxyy}$. 

All cases are representative (\textit{e.g.} see Appendix~\ref{sec:ensemle_average}), and in all cases the system finds a marginal absorbing state and stores memory. The details differ before the MAM is reached (black and orange data), but the final readout measurements (red data) tell the same story. The only anomaly is in Fig.~\ref{fig:memory}b(i), where $n_\mathrm{steps}$ does not show a kink at $\numin$ but remains small for small $\nuread$~\footnote{While we do not fully understand this, it is worth noting that $n_\mathrm{steps}$ is the weakest of our measurements since it depends on the details of the optimization algorithm, which are not controlled for.}.

Despite the relative similarity in the data, we do not claim that all systems converge to a MAM, and we do find exceptions. Most notably, we have not found an absorbing manifold when the training range includes very large Poisson's ratios. This is because soft spheres exhibit large $\nu$ at lower densities where the solid approaches marginal stability and structural rearrangements are unavoidable.

\subsection{The dynamics of contact changes}
We now turn our attention to understanding how a system reaches an absorbing manifold and why this manifold is marginally absorbing. 
\rev{These questions will be fully answered in Section~\ref{sec:mechanism}, but first, a significant clue comes from looking at the forming and breaking of particle-particle contacts. }
{First, note that large-scale structural rearrangements are not observed after the system reaches a MAM. Such rearrangements occur for earlier training cycles, e.g. Fig.~\ref{fig:methods}c, and presumably help steer the system away from minimally stable configurations, but they do not appear critical for the precise encoding of memory. However, particle-particle contacts can form and break without causing a structural rearrangement, e.g. Fig.~\ref{fig:methods}c, and these topological changes to the configuration will prove to be the key to understanding MAM formation.}

To study \rev{this}{the dynamics of contact changes}, we define the contact map $\mathcal{C}_{ij}(t)$ as 
\begin{equation}
    \mathcal{C}_{ij}(t) = \Theta\left(\sigma_{ij}(t) - r_{ij}(t)\right),
\end{equation}
where $\Theta$ is the Heaviside step function, so that $\mathcal{C}_{ij}(t)=1$ if particles $i$ and $j$ are in contact at optimization step $t$ and $\mathcal{C}_{ij}(t)=0$ otherwise. 

First, we compare the contact network before and after training from $\numax$ to $\nuread$ to compute the net number of contacts that change during training:
\begin{equation}
    n_\mathrm{cc}^\mathrm{net} = \sum_{i, j>i} \left| \mathcal{C}_{ij}(n_\mathrm{steps}) - \mathcal{C}_{ij}(0) \right|.
\end{equation}
Figure~\ref{fig:memory}(v) shows that after cyclic training, there are few, if any, contact changes between $\numax$ and $\nuread$ provided $\nuread$ is within the training range. Unlike the other quantities we have looked at, however, we do not observe consistent kinks or jumps at $\numin$ and $\numax$ that indicate memory. 

Interestingly, the small values of $n_\mathrm{cc}^\mathrm{net}$ do \textit{not} mean that contacts do not form and break repeatedly during optimization. To see this, we define the cumulative number of contact changes during optimization:
\begin{equation}
    n_\mathrm{cc}^\mathrm{cum} = \sum_{t=1}^{n_\mathrm{steps}} \sum_{i, j>i} \left| \mathcal{C}_{ij}(t) - \mathcal{C}_{ij}(t-1) \right|.
\end{equation}
Unlike $n_\mathrm{cc}^\mathrm{net}$, if a contact forms and then breaks, or vise versa, this does not cancel out in $n_\mathrm{cc}^\mathrm{cum}$. Figure~\ref{fig:memory}(vi) shows that $n_\mathrm{cc}^\mathrm{cum}=n_\mathrm{cc}^\mathrm{net}$ within the training range: on the rare occasion that a contact changes, this change persists during optimization. This is not true outside of the training range: in most cases, $n_\mathrm{cc}^\mathrm{cum}$ becomes much larger than $n_\mathrm{cc}^\mathrm{net}$, indicating a large number of contact changes that are then undone.  

This suggests that one of the consequences of reaching an absorbing manifold is that \textit{temporary} contact changes are suppressed along this manifold. For example, in Fig.~\ref{fig:methods}c, a single contact repeatedly forms and breaks, but this is not observed after cyclic training, Fig.~\ref{fig:methods}d. But why do contact changes matter? The answer is that the Poisson's ratio, and other elastic constants, depend on the Hessian matrix of second derivatives of the energy with respect to particle positions, see Appendix~\ref{sec:cijkl_calculation}. Importantly, our interaction potential, Eq.~\eqref{eq:hmorse}, is zero when particles do not overlap and proportional to their overlap raised to the power $2.5$ when they do. This means that the Hessian, and thus $\nu$, is continuous when contacts form or break. However, training depends on the gradient of $\nu$, not $\nu$ itself, and the gradient of $\nu$ is discontinuous when a contact changes because it depends on the third derivative of the energy. In the next section, we will see that these gradient discontinuities (GDs) are \textit{the} critical feature that drives the system into a MAM with encoded memory.

\begin{figure*}
    \centering
    \includegraphics[width=\linewidth]{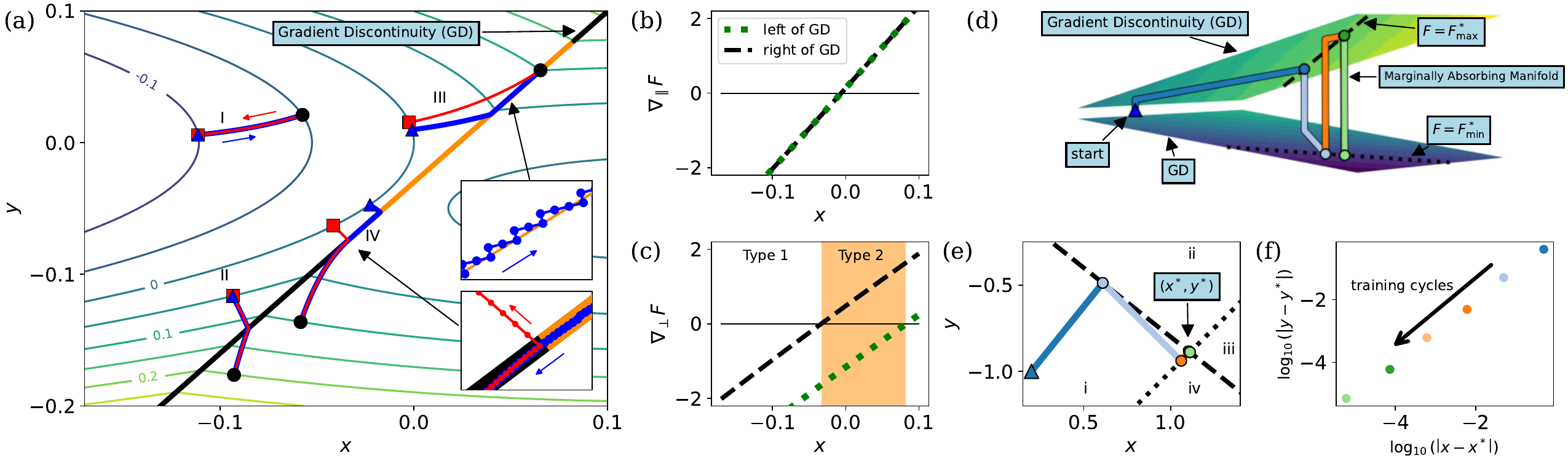}
    \caption{
    Gradient Discontinuity Learning.
    (a) An example training function $F(x,y)$ with a Gradient Discontinuity (GD) shown by the thick line (black when it is Type 1, orange when it is Type 2). Training cycles from four cases (labeled I - IV) begin at blue triangles, follow blue steepest ascent paths until a specified value of $F$ is reached (black circles), and then follow red steepest descent paths until $F$ returns to its original value (red squares). The insets zoom in on particular regions of interest.
    (b) and (c) give the parallel and perpendicular components, respectively, of $\nabla F$ just to the left and to the right of the GD, shown as a function of $x$. $\nabla_\parallel F$ must be continuous because $F$ is continuous. A Type 2 GD is defined as when $\nabla_\perp F$ changes sign at the GD, as indicated by the orange region.
    (d) An example training function of three variables where $F(x,y,z)=z$ between two planar Type 2 GDs. Starting at the blue triangle, cyclic training between $\Fmax$ and $\Fmin$ brings the system to a marginally stable manifold (MAM). The dashed lines show where values of $F$ at the GDs equal $\Fmax$ and $\Fmin$. 
    (e) A projection of the parameters during training onto the x-y plane. 
    (f) Cyclic training brings parameters iteratively closer to $(x^*,y^*)$\rev{, as discussed in the text}{}.
    }
    \label{fig:model2d}
\end{figure*}

\section{Gradient Discontinuity Learning\label{sec:mechanism}}
We now present a general mechanism called Gradient Discontinuity Learning that can lead to memory formation through cyclic training. We consider an arbitrary ``training function'' $F:\mathcal{R}^n \to \mathcal{R}$, that takes in $n$ parameters $\bm \theta$ and returns a scalar quantity that we wish to cyclically train. For clarity, we do not construct an objective function, but instead consider training dynamics that follow steepest ascent/descent paths in parameter space. 
We also assume that $F$ is continuous, but, crucially, we allow discontinuities in its gradient. As we will see, these Gradient Discontinuities (GDs) guide the exploration of the parameters under cyclic training, leading to the formation of a MAM that stores memory. 

While we will keep this discussion general in this section, one can think of $F$ as the Poisson's ratio, the parameters as the various particle diameters, and the GDs as deriving from contact changes. Here, we are not considering any effects analogous to structural rearrangements caused by instabilities in the packing, which lead to discontinuities in the value of $F$ rather than its gradient. 


\subsection{Gradient discontinuities guide parameter exploration}
We start with an illustrative example that demonstrates how GDs can lead to nontrivial parameter exploration under cyclic training. Consider the simple $n=2$ dimensional training function shown in Fig.~\ref{fig:model2d}a, which is given by
\begin{equation}
    F(x,y) = \tilde F(x,y) - f(x,y) \Theta(f(x,y)),
\end{equation}
where $\Theta$ is the Heaviside function, 
\begin{equation}
    \tilde F(x,y) = x + x^2 + 10y^2 + x^3,
\end{equation}
and
\begin{equation}
    f(x,y) = 1.3x - y - 0.03.
\end{equation}
$F(x,y)$ has a GD along the line given by $f(x,y)=0$ (shown by the solid black/orange line). Figure~\ref{fig:model2d}a also shows four example training cycles, labeled I-IV. Each case begins at an initial set of parameters (blue triangles) with function value $\Fmin$, follows the steepest ascent path (blue lines) until it reaches $\Fmax$ (black circles), and then follows the steepest descent path (red lines) back down until it returns to $\Fmin$ (red squares). 

Case I shows that when the ascent path does not cross the GD, the descent path simply returns the system to where it began, because the steepest descent direction is always the opposite of the steepest ascent direction~\footnote{In practice, the finite step size of any iterative ascent/descent algorithm will always result in some small amount of parameter drift.}.
In Case II, the ascent path crosses the GD, leading to a sharp and noticeable change in the direction of the path. Nevertheless, the descent path still reverses this, returning the system to where it began. In both of these cases, the system begins in an absorbing manifold because training is completely reversible. However, neither of these manifolds are marginally absorbing because they remain absorbing if the training range is widened. 

In contrast, Cases III and IV do not end where they begin, and there is a clear, nontrivial interaction with the GD. To understand what makes these cases different, Fig.~\ref{fig:model2d}b and c quantify the discontinuity in the gradient by showing $\nabla_\parallel F$ (b) and $\nabla_\perp F$ (c), the component of $\nabla F$ parallel and perpendicular to the GD, respectively, evaluated just to the left and just to the right of the GD. Since $F$ itself is continuous, $\nabla_\parallel F$ must be the same on both sides of the GD, as confirmed by Fig.~\ref{fig:model2d}b. However, Fig.~\ref{fig:model2d}c shows that there is a strong change in $\nabla_\perp F$ from one side to the other.

Now, when the sign of $\nabla_\perp F$ is the same on both sides of the GD, as it is in Case II, then a gradient pointing towards the GD on one side will point away from the GD on the other. This allows a gradient ascent/descent path to ``go through'' the GD, and we call this a Type 1 GD. However, the orange region in Fig.~\ref{fig:model2d}c shows where $\nabla_\perp F$ changes sign across the GD, and the gradient points towards the GD on both sides. Thus, whichever side you are on, gradient ascent paths will always bring you back towards the GD. We call this a Type 2 GD. In Fig.~\ref{fig:model2d}a, we color the GD line black to indicate where it is Type 1, and orange to indicate where it is Type 2.

The consequence of a Type 2 GD can be seen in Case III in Fig.~\ref{fig:model2d}a: when the blue ascent path hits the GD, it becomes bound to the discontinuity, oscillating back and forth between the two sides of the GD while simultaneously moving in the direction of $\nabla_\parallel F$ (up and to the right). This oscillation can be seen clearly in the first inset, where individual steps of the steepest ascent algorithm are shown by blue dots. The process stops when $F$ reaches the desired $\Fmax$ (in this case $\Fmax=0.1$) at the black dot. The key is that this oscillation about the Type 2 GD is not reversible. Starting from the black dot, the red steepest descent path is not bound to the GD but instead flows downhill away from it~\footnote{Note that the red descent path happens to start very slightly to the left of the GD, so it flows leftwards, but it could just as easily have started to the right of the GD and flowed in the opposite direction, depending on where in the oscillation the ascent path stops.}. By the time the system returns to $\Fmin$ (red square), the parameters have shifted along the $F=0$ contour line. The red path is an absorbing manifold, and continued training between $\Fmin=0$ and $\Fmax=0.1$ will return the system right up to the GD at the black dot. Furthermore, it is marginally absorbing because training even a little beyond $\Fmax$ will require further irreversible movement along the GD\rev{}{~\footnote{Note that $\Delta\Theta_\mathrm{RP}$ is the distance moved alone the $F=0$ contour, which increases continuously from zero when training is pushed beyond $\Fmax$. This explains the smooth increases in the readout measurements observed in Fig.~\ref{fig:memory}}}. Thus, the value of $\Fmax$ constitutes a memory as it can be retrieved through return-point measurements. 

Interestingly, a Type 2 GD can turn into a Type 1 GD, and Case IV shows that an ascent path bound to the GD will fall off and follow a ``normal'' path precisely where this transition occurs. 
This leads to a non-marginally absorbing manifold (red path) that intersects the GD precisely at the point where it transitions from Type 2 to Type 1. Note that even though the manifold is not marginally absorbing, intersecting the GD at this precise point can be viewed as a different type of memory, although we will not pursue this intriguing possibility further here.

\subsection{Multiple GDs can lead to MAMs with 2-sided memory}
Figure~\ref{fig:model2d}a demonstrates a mechanism for exploring functions of multiple variables through training, whereby steepest ascent/descent paths become bound to and move along Type 2 GDs. We now demonstrate how multiple GDs can lead to MAMs with 2-sided memory. 

Figure~\ref{fig:model2d}d shows two planar Type 2 GDs in a $n=3$ dimensional training function $F(x,y,z)$. $F(x,y,z)=z$ between the GDs, and the shading along the GDs indicates $F$. Starting between the two GDs at the blue triangle, the steepest ascent path (dark blue) first goes straight up until it reaches the top GD, and then moves along the GD until it reaches $\Fmax$ (dashed line). This is completely analogous to Case III in Fig.~\ref{fig:model2d}a. The system next follows the steepest descent path (light blue), which starts off going straight down until it reaches the bottom GD, and then moves along the GD until it reaches $\Fmin$ (dotted line). 

Crucially, the gradient $\nabla_\parallel F$ along the bottom GD is unrelated to $\nabla_\parallel F$ along the top GD, so the descent path does not reverse the ascent path. Figure~\ref{fig:model2d}e shows the projection of these paths onto the $x$-$y$ plane, making it clear that the ascent and descent paths affect $(x,y)$ in different ways. 
Now, a second ascent path up to $\Fmax$ (orange) will again hit and move along the top GD. Figure~\ref{fig:model2d}f shows that continued cyclic training brings the values of $x$ and $y$ iteratively closer to $x^*$ and $y^*$, 
defined by the intersection of the dashed and dotted lines in Fig.~\ref{fig:model2d}e. At this point, the ascent and descent paths only move vertically and end precisely at the GDs, meaning the system has approached a MAM with memory at both $\Fmin$ and $\Fmax$.

Figure~\ref{fig:model2d}e divides the $x-y$ plane into four quadrants. 
Cyclic training will result in a MAM with two-sided memory at $(x^*,y^*)$ whenever the initial parameters are in Quadrant~i (provided they are also between the two GDs). In Quadrant~iii, no memory is formed because the top and bottom GDs are above $\Fmax$ and below $\Fmin$, respectively, so the ascent/descent paths do not hit the GDs (as in Case I in Fig.~\ref{fig:model2d}a). In Quadrant~ii and iv, one-sided memory at either the dashed or dotted line is formed in a single training cycle (as in Case III).

\subsection{Gradient Discontinuity Learning in practice \label{sec:training_algorithms}}
So far in this section we have used simple examples to show how GDs can guide the evolution of parameters during cyclic training. In real systems with $n$ parameters, GDs are (up to) $n-1$ dimensional surfaces~\footnote{Lower dimensional GDs are of course possible, but they are probably much less important due to the smaller chance of encountering them. They do not appear in our sphere packings, and we do not consider them further.}. 
When a steepest ascent/descent path encounters a Type 2 GD, it binds to the GD and follows the steepest ascent/descent path given by $\nabla_\parallel F$, falling off if and when the GD becomes Type 1. 
We call this process Gradient Discontinuity Learning (GDL). 
Nontrivial learning occurs when cyclic training encounters multiple GDs, and the ascent/descent paths become much more complicated for higher-dimensional training functions and with curved and intersecting GDs.

We have presented these ideas by asserting that training dynamics follows steepest ascent/descent paths until a desired function value is reached. In practice, however, it is more common to minimize an objective function, such as $\ell = (F-F^*)^2$. Clearly, discontinuities in $\bnabla_\btheta F$ lead to discontinuities in $\bnabla_\btheta \ell = 2(F-F^*)\bnabla_\btheta F$, and the theory is unaffected. Interestingly, many inverse design problems include constraints on the parameters. One way to incorporate such constraints into the objective is to define a function $c(\btheta)$ that is positive when the constraint is active,
and define the objective as
\begin{equation} \label{eq:loss_with_constraints}
    \ell = (F-F^*)^2 + a c(\btheta) \Theta(c(\btheta)),
\end{equation}
where $\Theta$ is again the Heaviside step function and $a$ is a constant. Now, the gradient of $\ell$ becomes
\begin{equation}
    \bnabla_\btheta \ell = 2 (F-F^*)\bnabla_\btheta F + a [\bnabla_\btheta c(\btheta) ] \Theta(c(\btheta)),
\end{equation}
meaning that as long as $\left.\bnabla_\btheta c \right|_{c=0}\neq 0$, there is a discontinuity in the gradient of $\ell$ when the constraint becomes active. Furthermore, this GD is always a Type 2 GD in the limit of hard constraints, $\left|a\right|\to\infty$.
Therefore, even if $\nabla_\btheta F$ is continuous, GDL can nevertheless exploit constraints to encode memory.

Learning without discontinuities in $\bnabla_\btheta F$ can also result from the practical details of optimization.
Perfectly reversible, non-absorbing manifolds like Case I in Fig.~\ref{fig:model2d}a rely on training dynamics following precise ascent/descent paths. However, training dynamics rarely follows such paths exactly. 
Finite step sizes always lead to small deviations from ideal paths, and many algorithms (\textit{e.g.} \rev{congugate}{conjugate} gradient, LBFGS, or any momentum-based algorithm) intentionally move in directions informed by, but not parallel to, the gradient. 
The consequence of this is that GDs are not strictly necessary\rev{ for learning}{}: regions of high curvature can cause actual training paths to be non-absorbing, causing nontrivial movement in the parameters \rev{}{ in directions perpendicular to the gradient}. \rev{We will see an example of this in our data momentarily.}{}
GDL should be viewed as a precise description of the ideal case, where \rev{learning is guided directly by GDs, but the ideas developed by GDL can extend more broadly.}{parameter exploration in directions perpendicular to the gradient is guided by GDs.}
\rev{We leave a proper exploration of learning without any GDs to future work.}
{While other mechanisms for such parameter exploration exist, \textit{e.g.} Dillavou et al.~\cite{dillavou2025understanding} show that hardware imperfections in experimental learning networks introduce deterministic biases into the update rule that drive parameter drift during repeated training, GDL provides a simple mechanistic understanding of memory formation.}

\rev{}{Finally, can GDL apply to systems, like sheared solids and suspensions, that are trained through an applied field rather than through explicit inverse design?
We can generalize GDL by considering any system characterized by parameters $\btheta$ subject to environmental variables $e$, where the parameters respond to the environmental variables so as to maintain a constraint $G(e, \btheta)=0$ imposed by the physics of the system. 
This constraint could, for example, specify a desired target property (\textit{i.e.} $G(F^*,\btheta)=F(\btheta)-F^*=0$)
or impose force balance on all particles in a sheared suspension, see below. 
The solutions to this constraint define an implicit function $\btheta^*(e)$, 
so that for any $e$, $G(e, \btheta^*(e))=0$. 
When $e$ changes, the parameters will change in directions determined in part by $\bnabla_\btheta G$, so that the parameter trajectories are reversible provided that the gradients of $G$ are continuous. When the system encounters gradient discontinuities, irreversible dynamics will move the parameters along the gradient discontinuities. GDL therefore provides a straightforward mechanism for moving the parameters in directions perpendicular to $\bnabla_\btheta G$, though we leave a more thorough investigation of this scenario to future work.
}

\subsection{GDL explains the observed MAMs in disordered packings} 

Does GDL actually explain the results of Figs.~1-3? As already stated, GDs arise from particles coming in and out of contact, and typical training encounters many such events, so GDs have the opportunity to influence the training. Furthermore, our mechanism makes precise predictions that we can test. If a MAM with 2-sided memory is reached through this GD-based mechanism by cyclically training between $\numin$ and $\numax$, then:
\begin{enumerate}
    \item[P1:] If any contacts change in the middle of the MAM (\textit{i.e.} between $\numin$ or $\numax$), they must correspond to Type 1 GDs. 
    \item[P2:] At both ends of the MAM (\textit{i.e.} at $\numin$ and $\numax$), a contact between two particles should be just on the verge of forming or breaking, with the corresponding GD being Type 2. If the MAM only has 1-sided memory, then this applies only to that end.
    \item[P3:] During the learning process, there should be times of rapid contact-change oscillations, corresponding to the system moving along a Type 2 GD.
\end{enumerate}
These predictions are not trivial: they are unlikely to occur by chance and could easily be wrong if memory was formed through a different mechanism. Unlike Prediction P1, Predictions P2 and P3 assume that learning is guided precisely by a GD rather than a region of high curvature (see discussion in Sec.~\ref{sec:training_algorithms}), and therefore may not always hold. This is consistent with our observations: P1 and P3 always hold while P2 is true most of the time, but with a few exceptions.  


\begin{figure}
    \centering
    \includegraphics[width=\linewidth]{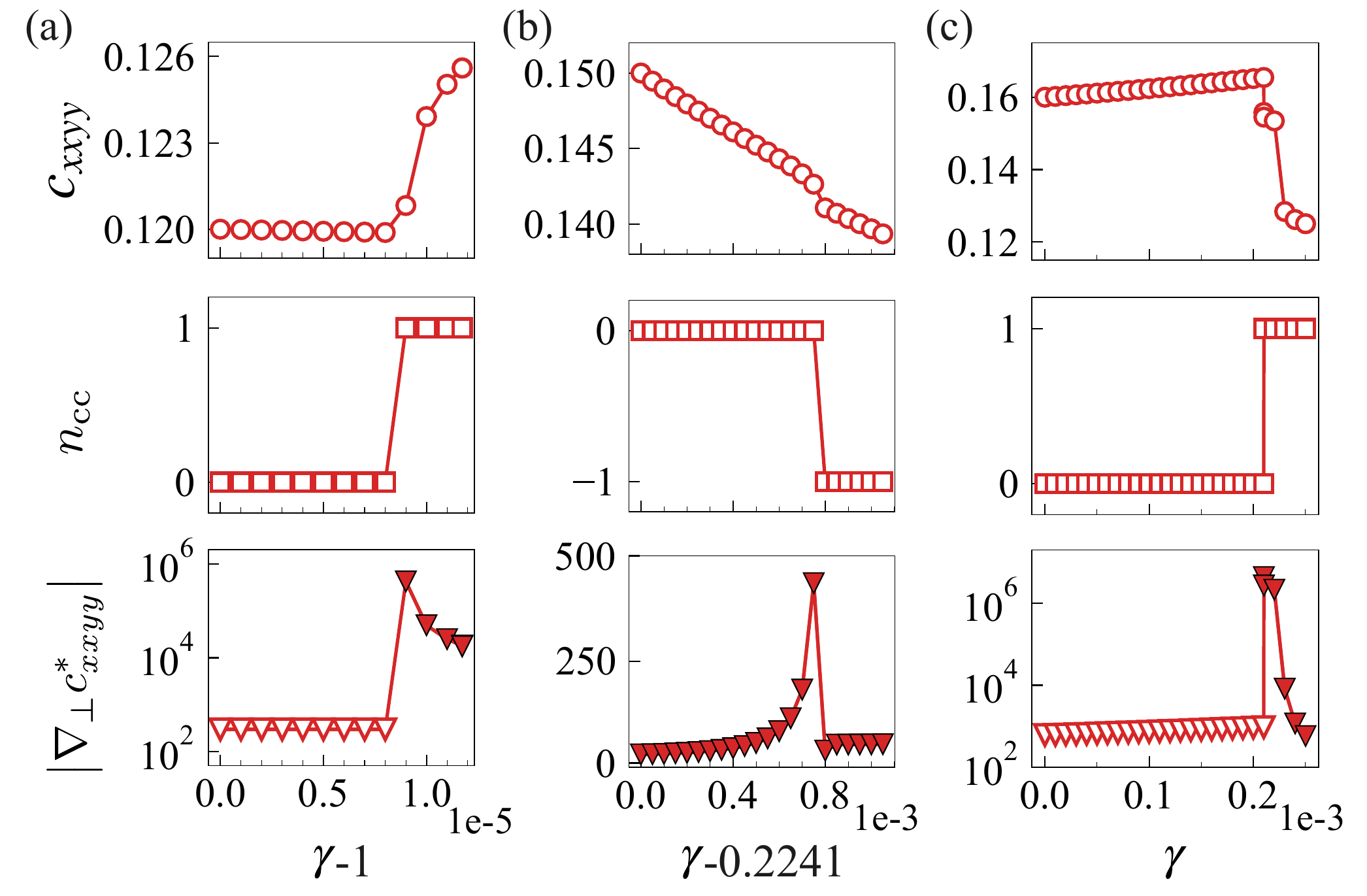}
    \caption{
    Confirming predictions made by Gradient Discontinuity Learning.
    Using the system from Fig.~\ref{fig:memory}d after cyclic training, we systematically set the parameters to be along the line defined by $\boldsymbol{\vartheta}$ (Eq.~\eqref{eq:vartheta}) to identify contact changes and probe the GD type. 
    Specifically, we set $\btheta=\btheta_{c^*_\mathrm{xxyy,max}} + \gamma \boldsymbol{\vartheta}$, so that $\btheta = \btheta_{c^*_\mathrm{xxyy,max}}$ at $\gamma=0$ and $\btheta = \btheta_{c^*_\mathrm{xxyy,min}}$ at $\gamma=1$. 
    Data is shown as a function of $\gamma$ and is focused around three contact changes: (a) just below the training range, near $\gamma=1$, (b) in the middle of the training range, near $\gamma=0.2241$, and (c) just above the training range, near $\gamma=0$. The top row shows the value of $c_\mathrm{xxyy}$, the middle row shows the change in the number of contacts, allowing us to identify the moment of contact change, and the bottom row shows the magnitude $\left|\nabla_\perp c_\mathrm{xxyy}\right|$, with positive (negative) values of $\nabla_\perp c_\mathrm{xxyy}$ given by open (closed) symbols. 
    As predicted, the GD in the middle of the MAM (b) is of Type 1 because $\nabla_\perp c_\mathrm{xxyy}$ does not change sign (Prediction P1), while the two GDs at the edges of the MAM are Type 2 because $\nabla_\perp c_\mathrm{xxyy}$ does change sign (Prediction P2). 
    }
    \label{fig:predictions}
\end{figure}

We have already seen the oscillations of Prediction~P3 in the inset to Fig.~\ref{fig:methods}c, which is representative and generic. Predictions~P1 and P2 are demonstrated in Fig.~\ref{fig:predictions}, which shows contact changes and $\nabla_\perp c_{xxyy}$ at various points of the MAM from Fig.~\ref{fig:memory}d. Specifically, we zoom in around 3 contact changes (shown by the middle row) that occur at the beginning (a), middle (b), and end (c) of the MAM. a) and c) show that contacts change almost immediately outside of the MAM, and that $\nabla_\perp c_{xxyy}$ changes sign (Prediction~P2). Conversely, this MAM happens to have a contact change in the middle, but here $\nabla_\perp c_{xxyy}$ does not change sign (Prediction~P1). The counterexamples for Prediction~P2 are the systems in Fig.~\ref{fig:memory}b and c, where no contacts are on the verge of changing near $\numax$. 
These nontrivial observations are consistent with what we would expect from GDL, and the counterexamples motivate future investigations into the more complicated scenario of learning without GDs.




\section{Discussion \label{sec:discussion}}
\rev{We have shown that memory can emerge in tunable disordered solids through cyclic training, reminiscent of cyclically sheared suspensions or dense solids
. After iteratively training a given observable $F$ (\textit{e.g.} the Poisson's ratio) between two values, $\Fmin$ and $\Fmax$, we observe the emergence of memory through various return-point measurements, the ``ease'' of training, and the direction of parameter changes. This memory is associated with the system approaching a marginally absorbing manifold (MAM), where training returns the system to the same state after a full cycle. We also present a general mechanism for the formation of memory, called Gradient Discontinuity Learning (GDL), where training is guided by discontinuities in the gradient of $F$ with respect to the tunable parameters (\textit{e.g.} particle diameters). These Gradient Discontinuities (GDs) impact steepest ascent/descent paths, allowing us to predict and explain much of the behavior we observe. }{We have shown that cyclic training in tunable disordered solids can produce physical memory by driving the system toward a marginally absorbing manifold (MAM) that encodes the training range. This memory is highly reminiscent of that observed in cyclically sheared suspensions but is distinct from the return-point memory seen in cyclically sheared disordered solids, which is characterized by periodic but not reversible dynamics~\cite{RevModPhys.91.035002, paulsen2025mechanical-d8c, mungan2025self-organization-787}.  
We further proposed a general mechanism, Gradient Discontinuity Learning (GDL), in which discontinuities in the gradient of the training function guide the training dynamics and help generate MAMs.} 

These results bring us closer to a general understanding of the effects of a slowly changing environment on a system undergoing optimization. While we exclusively consider changes to the desired value of an observable, many environmental variables instead affect observables directly. \rev{}{In our case, for example, we could keep the desired Poisson's ratio fixed while cyclically varying the total packing fraction, which would systematically affect the optimized solution in much the same way as changing the target behavior.
} GDL theory \rev{applies in these cases as well}{can apply here as well (see Sec.~\ref{sec:training_algorithms})} because changing such variables will cause the optimal solution to change, and if this optimal solution encounters one or more Type 2 GDs, the system may approach a MAM. 

\rev{Importantly, reaching a MAM is never guaranteed -- in our athermal sphere packings, it is somewhat surprising that it is even possible to train between $\nu=0.5$ and $\nu=0$ without structural rearrangements. Nevertheless, GDs provide a concrete and generic mechanism for guiding the optimization parameters, creating the potential for memory. }{
While GDL provides a mechanism for memory formation, there are no guarantees that a given system will actually reach a MAM. Even if there are GDs in the training function, if their density is too low then training may not encounter them, resulting in an absorbing manifold that is not marginal and thus does not store memory (as in Case I in Fig.~\ref{fig:model2d}a). On the other hand, if the density of GDs is too high, then training may not be able to find an absorbing manifold that spans the training range. In our athermal sphere packings, it is somewhat surprising that it is possible to train between $\nu=0.5$ and $\nu=0$ without contact changes, and we do not find MAMs when training to higher $\nu$'s. 
}

In this paper, we have focused on training a single observable. However, just as one can form memory from cyclically shearing in multiple directions~\cite{adhikari2025encoding-cfe,lindeman2025multidimensional-421}, one could consider training multiple properties. Bulk mechanical properties of disordered solids are known to be fully independent~\cite{zu2025fully-43b}, meaning that they can be tuned separately and even simultaneously, opening the door for a variety of interesting training protocols. As a simple preliminary demonstration of this, Fig.~\ref{fig:appendix4} in Appendix~\ref{sec:multiple_features} shows that we can approach an absorbing manifold by alternately training between two different elastic constants (\textit{i.e.} first a cycle of one and then a cycle of the other). However, we leave a proper quantification and analysis of the memory contained in such systems to future work. 

One major consideration that we have not discussed is the role of noise. Keim and Nagel~\cite{PhysRevLett.107.010603} showed that \rev{}{in sheared suspensions, } thermal noise plays a critical role in the formation of \rev{memory in sheared systems, allowing for the possibility of multiple memories}{multiple transient memories}, and it is not clear if GDL can lead to multiple memories in the same way. As discussed above, training does not exactly follow steepest ascent/descent paths, which could be interpreted as a form of noise, but it would be interesting to systematically investigate the consequence of introducing random noise either to the initial parameters at the beginning of each optimization or during every optimization step. 
\rev{}{It would also be interesting to study what happens if the system is only allowed to partially relax during the training process: rather than fully minimizing the energy of the configuration during each training step, one could update parameters on the same time scale as the system relaxes, potentially achieving faster and more robust memory.}

\rev{}{
Our results are closely related to, but distinct from, the oscillatory training framework of Falk et al.~\cite{falk2023learning}, who showed that switching between two incompatible target functionalities selects for adaptable regions of design space where the two solutions are close together in parameter space. Both approaches demonstrate that cyclic training produces nontrivial emergent structure beyond simply achieving the current target, and both rely critically on the degeneracy of high-dimensional design spaces to make this possible. 
While Falk et al. frame this emergent structure in terms of adaptability, we find that cyclic training also encodes memory of the training range in the form of a MAM. These two perspectives are complementary: an adaptable system in the sense of Falk et al. has implicitly found a region of parameter space that is low-cost to traverse between functions, which is related to our finding that the MAM makes training within the range dramatically easier. We further propose GDL as a potential mechanism underlying such behavior, though establishing this connection more rigorously, particularly for training protocols that differ from gradient descent, is an interesting direction for future work.}

While we have focused on training the elastic properties of disordered sphere packings, GDL is much more general and should be applicable in a wide range of settings. First and foremost, our results encourage cyclic training in all manner of systems that can be inverse-designed, especially those with clear GDs or complex constraints. However, GDL may also play a role even in systems that are not explicitly inverse-designed. We now speculate briefly about a few such possibilities.

\paragraph*{Sheared suspensions.}
\rev{We have already discussed the numerous work showing that cyclically sheared suspensions lead to return-point memory.
While the physics of these systems can be understood on their own, they can also be viewed through the lens of our theory. 
The act of shearing has the effect of changing the average affine shear strain, subject to the constraints that particles do not overlap. This can be mapped to Eq.~\eqref{eq:loss_with_constraints}, where the parameters are all particle positions, $F$ is the average affine shear strain, and there is a separate constraint for every pair of particles that provide the GDs necessary for learning.}{
The phenomenology we observe in trained disordered solids is very similar to that of cyclically sheared suspensions~\cite{pine2005chaos-3f0,Corté2007,menon2009universality-967,PhysRevLett.107.010603,keim2013multiple-1d6,pham2015particle-ebf,PhysRevLett.113.068301}.
While the physics of these systems can be understood on their own, they can also be viewed through the lens of our theory. 
Since the net force on every particle, which must vanish at all times, depends on the strain $\gamma$ and the non-affine position of every particle, $R_\mathrm{na}$, we have $F_\mathrm{net}(\gamma,R_\mathrm{na})=0$. This maps exactly to GDL, as discussed at the end of Sec.~\ref{sec:training_algorithms}, with GDs deriving from particle contacts.}

\paragraph*{\rev{}{Sheared jammed solids.}}
\rev{}{This presumably also can be done for memory formation in sheared jammed solids, but this situation is more complex with a very different form of memory~\cite{RevModPhys.91.035002}.
First, as observed by Arceri et al.~\cite{PhysRevE.104.044907}, marginal stability is crucial for memory formation, implying the presence of structural instabilities within the training range. 
Structural instabilities can lead to both irreversible or reversible rearrangements~\cite{regev2021topology-1b4}, and it is well known that such systems can develop memory while undergoing reversible rearrangements~\cite{adhikari2018memory-84e, PhysRevResearch.2.012004,paulsen2025mechanical-d8c}. However, this implies a discontinuity in the training function, not just in its gradient, and further work is needed to incorporate reversible rearrangements, discontinuities, and hysteretic cycles into the paradigm presented here. 
}

\paragraph*{\rev{}{Crumpled sheets.}}
\rev{}{
Another well-established form of physical memory is when a system can recall the maximum value of a previously applied perturbation. First observations include the Kaiser effect, describing the acoustic emission of a metal under strain~\cite{kaiser1964investigation}, and the Mullins effect, describing stretched rubber~\cite{mullins1948effect-d24}. This is also seen in crumpled sheets compressed by a piston of mass $m$~\cite{matan2001crumpling-c0f, RevModPhys.91.035002, PhysRevLett.118.085501}. The height of the piston is a reproducible function of $m$, provided that $m$ does not exceed the previous maximum applied mass, effectively encoding this previous maximum mass. Conceptually, this is very similar to the ``single-shot'' memory discussed in Case III in Fig.~\ref{fig:model2d}a, and the constraint that the crumpled sheet is internally stable presumably leads to discontinuities caused by further crumpling. While the application of GDL to crumpled sheets is therefore quite plausible, further work is again need to solidify this connection.
}

\paragraph*{Evolution.} 
\rev{}{The role of environmental variation in shaping adaptive behavior has been studied in population genetics models, where temporal correlations in environmental fluctuations have been shown to favor phenotypic switching strategies that exploit memory of past states~\cite{PhysRevE.96.032412}. This motivates the search for GDL-like mechanisms in biological evolution.}
\rev{T}{Importantly, t}he genotype-phenotype (GP) map between an organism's genotypes and phenotypes is highly degenerate~\cite{alberch1991genes,cowperthwaite2007mutational,pigliucci2010genotype,ahnert2017structural}, and genotypes can combine in highly nontrivial, collective ways to affect fitness~\cite{scheiner1993genetics,begun2007population,orr2009fitness,keren2016massively,chen2023gene}. Effective GDs can arise, \textit{e.g.}, whenever a phenotype is limited by multiple genotypes, or if a genotype can be deleterious past a threshold that depends on other genotypes. Furthermore, environmental changes are well known to play a key role in evolution, with comparable timescales for evolutionary and ecological change~\cite{carroll2007evolution-5c4}.
With all the necessary ingredients in place, it is at least plausible that GDL has played a role in the exploration of the apparent degeneracy in the GP map.

\paragraph*{Phenotypic plasticity and (re)adaptation.} 
Phenotypic plasticity, where a single genotype can exhibit multiple phenotypes depending on environmental inputs, may provide more direct examples than pure genetic evolution. 
For instance, the number of nuclei in muscle cells increases with muscle mass, but then remains roughly constant during muscle atrophy~\cite{staron1991strength-be1, schwartz2013muscle-00a}. This makes it significantly easier to gain muscle mass the second time, but only back to previous levels, creating a \rev{return-point }{}memory familiar to anyone who frequents the gym. Here, the training function (muscle mass) is constrained by the number of nuclei, creating a GD that guides the initial increase in nuclei. However, the easiest way to decrease muscle mass does not involve removing nuclei, creating a scenario similar to Case III in Fig.~\ref{fig:model2d}a.

As a second example that mixes phenotypic plasticity with genetic evolution, Ho \textit{et. al}~\cite{ho2020phenotypic-92a} show that ``organisms generally `remember' their ancestral environments via phenotypic plasticity,'' and present a potential mechanism for long-term memories. We can reformulate this mechanism in the context of GDL, as follows. The amount of expression of a hypothetical gene can be regulated through transcription factors that are controlled, in part, by environmental \rev{queues}{cues}. However, such phenotypic plasticity is limited by regulatory motifs, which are genetically controlled. This creates a discontinuity in the fitness gradient when transcription factors and regulatory motifs are in balance, with $\nabla_\parallel F$ projecting onto the regulatory motifs. Therefore, new environmental pressures requiring an increase in gene expression would require evolutionary adaptation to increase regulatory motifs. However, the decrease in gene expression required by a subsequent return to the original environment can be accommodated purely by phenotypic plasticity through a decrease in transcription factors. Therefore, one cycle of environmental change leads to a MAM with 1-sided memory.



There are many other situations where GDL may apply but where identifying the GDs would require further investigation. Clearly, both biological and artificial neural networks are possibilities. For example, it is possible that the topological connections of neurons in the brain encode history of past experiences, making it easier to, for example, relearn a forgotten language or skill. GDL may also play a role in mimetic evolution, or the formation of cultural, social, and political norms. 
While these examples are highly speculative, they highlight the potential for a far-reaching understanding of memory and learning.

Finally, it is worth putting our results in the context of the growing initiative exploring learning as a unifying concept across fields, which is the subject of this Special Collection. Specifically, it is important to clearly distinguish between the solving of an individual inverse problem, which we refer to as training or optimizing, from the system approaching a MAM and the emergence of memory. While the process of optimization can vary dramatically in different systems, GDL is agnostic to these details provided they have the effect of approximating some form of gradient descent. In this paper, we reserve the word learning to refer to the emergence of memory after cyclic training. A key distinction between this and many other phenomena associated with the word ``learning'' is that here the system was never asked to learn. Our learned behavior is not associated with an objective, either explicit or implicit, there is no reward, no bias, and no local rules are intelligently set up to guide a predetermined behavior. 
Whether or not this distinction is important will surely be decided in the coming years. 

\acknowledgements
We thank Nathan Keim, Aayush Desai, Nicholas Barton, and Gašper Tkačik for important and stimulating discussions. The work was funded by the Institute of Science and Technology Austria.

\appendix
\section{Calculation of elastic properties\label{sec:cijkl_calculation}}
The linear elastic properties of a system can be represented by the elastic modulus tensor \rev{$C_{ijkl}$}{$c_{ijkl}$}, which describes the second order terms of the Taylor expansion of the energy with respect to external bulk deformations,
\begin{equation*}
    \frac{\Delta U}{V^0}=\sigma^0_{ij}\epsilon_{ji}+\frac{1}{2}{\rev{C_{ijkl}}{c_{ijkl}}}\epsilon_{ij}\epsilon_{kl}+O(\epsilon^3)
\end{equation*}
where $\sigma^0$ gives the residual stresses of the initial state, $\epsilon$ is the strain tensor that describes applied boundary deformations, $V^0$ is the volume of the initial state, and $\Delta U$ is the change in the potential energy of the system.

If a small strain $\epsilon=\gamma\tilde{\epsilon}$ is applied, for some fixed $\tilde{\epsilon}$, the change in the energy is,
\begin{equation*}
    \frac{\Delta U}{V^0}=\tilde{\sigma}^0\gamma + \frac{1}{2V^0} \left(\frac{\partial^2U}{\partial \gamma^2} - \Xi^T(H^0)^{-1}\Xi\right)\gamma^2+ O(\gamma^3)
\end{equation*}
where $\tilde{\sigma}^0=\sigma^0_{ij}\tilde{\epsilon}_{ji}$, $\Xi=-\frac{\partial^2U}{\partial\gamma\partial R}$, and $H^0=\frac{\partial^2 U}{\partial R^2}$ is the Hessian matrix of the energy with respect to particle positions $R$. This allows one to determine the elements of $c_{ijkl}$ by calculating $\frac{1}{2V^0} \left(\frac{\partial^2U}{\partial \gamma^2} - \Xi^T(H^0)^{-1}\Xi\right)$ for different $\tilde \epsilon$. We use the implementation of this found in the \texttt{JAXMD} library~\cite{jaxmd2020}, further details and documentation can be found there. 

\rev{}{More common elastic constants like the bulk modulus, shear modulus, and the Poisson's ratio are simple functions of the elements of $c_{ijkl}$~\cite{Goodrich2014iu}. Since any finite disordered system is technically anisotropic, even if it is statistically isotropic in the thermodynamic limit, its elastic response depends on the direction of applied strain. In $d$ dimensions, there are $\frac{d(d+1)}{2}-1$ independent shear moduli. Thus, for finite systems, we use the so-called angle-averaged Poisson's ratio, $\nu=\frac{d-2\frac{\bar{G}}{B}}{2(d-1)\frac{\bar{G}}{B}+d}$, where $\bar{G}$ is the angle-averaged shear modulus,
\begin{equation*}
    \bar{G}=\frac{1}{\pi}\int_{0}^{\pi}G(\theta)d\theta, 
\end{equation*}
where $G(\theta)$ is the shear modulus corresponding to a shear applied at angle $\theta$. Details of this construction are discussed in~\cite{Goodrich2014iu}. 
The bulk modulus is calculated in terms of the elastic constants $c_{ijkl}$, $B=\frac{1}{d^2}\sum_{i,j}c_{iijj}$. 
}

\section{Gradient-based optimization algorithm \label{sec:optimization_algorithm}}
Once an energy-minimized configuration is reached for a given initial set of parameters $\btheta$, the Poisson’s ratio $\nu(\btheta)$ is calculated via linear response. Given a target Poisson’s ratio \(\nu^*\), the objective function is defined as the squared deviation, $l(\btheta) = (\nu(\btheta) - \nu^*)^2$. Thus, our aim is to decrease $l(\btheta)$ to zero using gradient descent algorithms.

The energy minimization, as well as the calculation of $\nu(\btheta)$ and $l(\btheta)$, are implemented in Python using the \texttt{JAXMD} library~\cite{jaxmd2020}, which is built upon \texttt{JAX}~\cite{jax2018github} and supports automatic differentiation. This framework allows for end-to-end differentiability of the entire calculations, allowing direct calculation of the gradient $\nabla_{\btheta}l$. 

To accurately and efficiently propagate gradients through the energy minimization step, we employ implicit differentiation, implemented via the \texttt{JAXOPT} library~\cite{blondel2022efficient}. This approach mitigates memory overhead typically associated with unrolled gradient computation. The calculated gradients \(\nabla_\btheta l\) are then used to update the parameters $\btheta$ using the RMSProp algorithm, as implemented in the \texttt{OPTAX} library~\cite{deepmind2020jax}. To enhance optimization efficiency, we incorporate a meta-learning scheme that dynamically adjusts the learning rate throughout training. The base learning rate is initialized at \(lr_0 = 1 \times 10^{-4}\), with a meta-learning rate of \(lr_m = 3 \times 10^{-4}\). 

At each iteration, the final configuration from the previous step is used as the starting configuration for the subsequent energy minimization, thereby maintaining consistency with the given energy minimum throughout training.

\section{Cyclic training protocol\label{sec:cyclic_training_appendix}}
As an example, we implement cyclic training between a maximum target Poisson’s ratio of \(\nu^*_{\mathrm{max}} = 0.5\) and a minimum of \(\nu^*_{\mathrm{min}} = 0.0\). The maximum number of optimization steps is set to 3000. The initial configuration is generated by placing particles randomly in the simulation box and quenching to a local energy minimum, which serves as the input state \(\bR(\btheta)\).

Cyclic training begins with a target value of \(\nu^*=\numax = 0.5\) and proceeds through a sequence of decreasing and then increasing target values. The procedure is outlined below:

\begin{itemize}
    \item[\textbf{Step 1.}] Train the input state \(\bR(\btheta)\) with the target Poisson’s ratio \(\numax = 0.5\), using the optimization algorithm described in Appendix~\ref{sec:optimization_algorithm}. This yields the final parameters \(\btheta_f\) and the corresponding configuration \(\bR\).
    
    \item[\textbf{Step 2.}] Using the output from Step 1 as input, repeat the optimization with a decreased target value, \(\nu^* = 0.4\).
    
    \item[\textbf{Step 3.}] Continue decreasing the target value in steps (e.g., \(\nu^* = 0.3, 0.2\), etc.), repeating the optimization process until \(\nu^*=\numin = 0.0\) is reached.
    
    \item[\textbf{Step 4.}] Reverse the direction by incrementally increasing the target value from \(\nu^* = 0.1\) back to \(\nu^*=\numax = 0.5\), using the final configuration from each step as the initial configuration for the next.
    
    \item[\textbf{Step 5.}] Evaluate the objective function curves for every optimization in the cycle. If all final objective function values remain below a predefined threshold and all curves are smooth and continuous (i.e., without discontinuities or spikes), the cyclic training is deemed successful. Otherwise, repeat Steps 1--4 until either a successful cycle is obtained or the number of cycles exceeds 30. 
\end{itemize}

After successful termination, an additional refinement cycle is performed in which the learning rate is fixed at \(10^{-6}\) for each optimization step to minimize the influence of learning rate adaptation.

For the readout cycle, we run a separate sequence with fixed learning rate \(lr = 10^{-6}\), following the path \(\numax = 0.5 \rightarrow \nu^*_{\mathrm{read}} \rightarrow \numax = 0.5\), starting from the final configuration and parameters obtained at the end of certain training cycles. For comparison with the cyclic training, we perform a successful single-target training with a target \(\numax = 0.5\), starting from the initial state and parameters. The resulting optimized configuration and parameters are used as the input for the readout cycle, which we defined as ``cycle=0'' in Fig~\ref{fig:memory}.

\section{\rev{}{The role of the loss threshold and learning rate on memory formation} \label{sec:analysis_nonzero_changes}}
\begin{figure}
    \centering
    \includegraphics[width=0.7\linewidth]{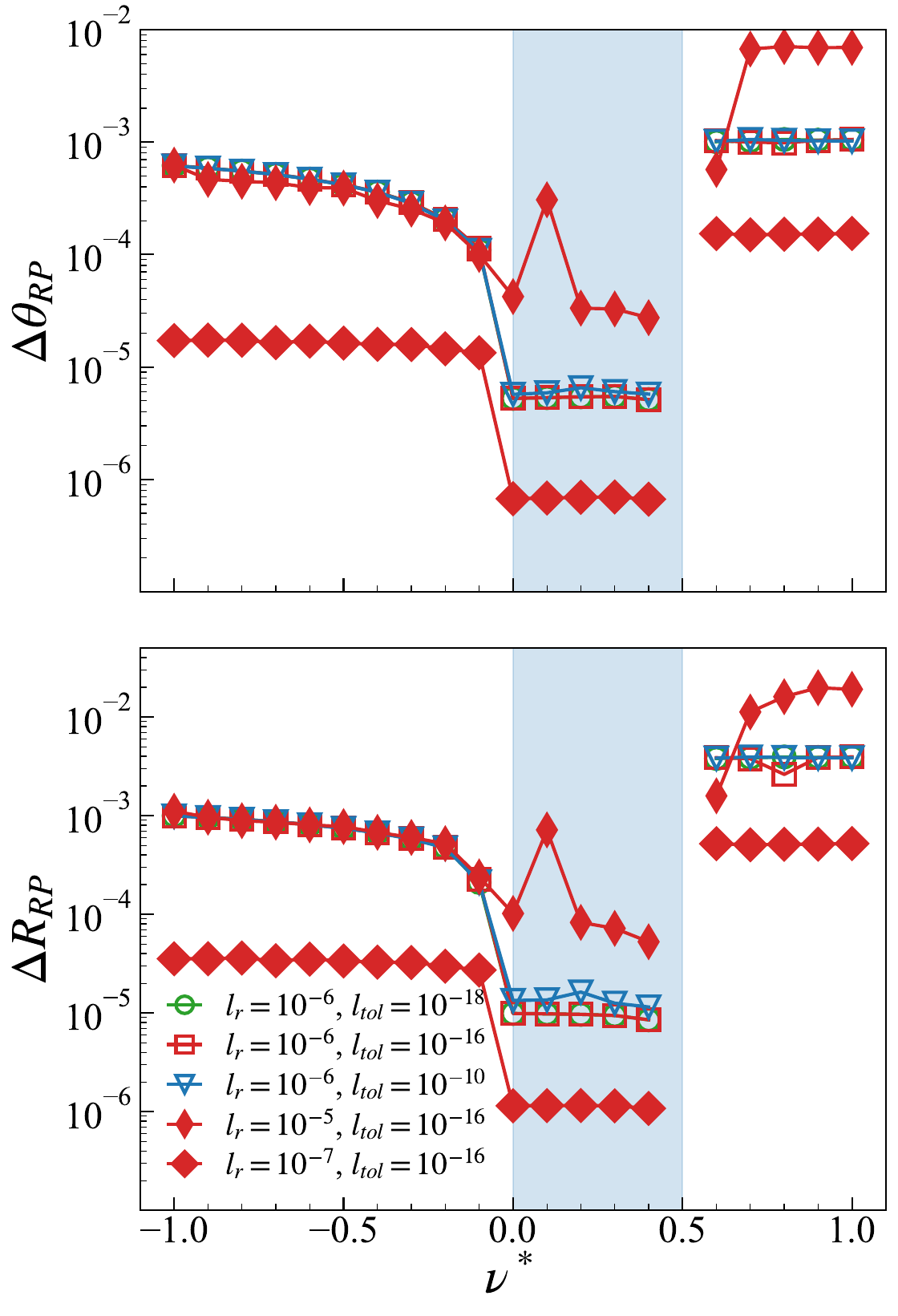}
    \caption{Robustness of the observed memory. Retraining results for the example shown in Fig.~\ref{fig:memory}a, with varying objective tolerances $l_{tol}$ and learning rates $l_r$ spanning several orders of magnitude. With changing $l_{tol}$, $\Delta\theta_{RP}$ and $\Delta R_{RP}$ after retraining remain unchanged, indicating that the observed memory is not due to incomplete convergence at the chosen cutoff. However, reducing $l_r$ decreases the magnitude of the deviations, demonstrating that they arise from numerical drift in the optimization dynamics.}
    \label{fig:appendix5}
\end{figure}

\rev{}{To determine whether the memory observed in the main text depends on numerical choices in the optimization procedure, we systematically varied both the objective tolerance $l_{tol} $ used to terminate retraining and the learning rate. We focus on the example shown in Fig.~\ref{fig:memory}a, with the results presented in Fig.~\ref{fig:appendix5}. In the main text, retraining or readout was stopped once the objective function fell below $10^{-16}$. We find that varying this threshold over several orders of magnitude does not affect the magnitude of the deviations in parameters or particle positions observed in Fig.~\ref{fig:memory}a. This demonstrates that the behavior is not a consequence of incomplete retraining due to the chosen cutoff.}

\rev{}{In contrast, the magnitude of the deviations decreases with decreasing the learning rate $l_r$, as shown in Fig.~\ref{fig:appendix5}, indicating that they arise from finite step size effects in the optimization dynamics. Note that reducing the learning rate does not qualitatively affect the existence of the MAM.}

\begin{figure}
    \centering
    \includegraphics[width=\linewidth]{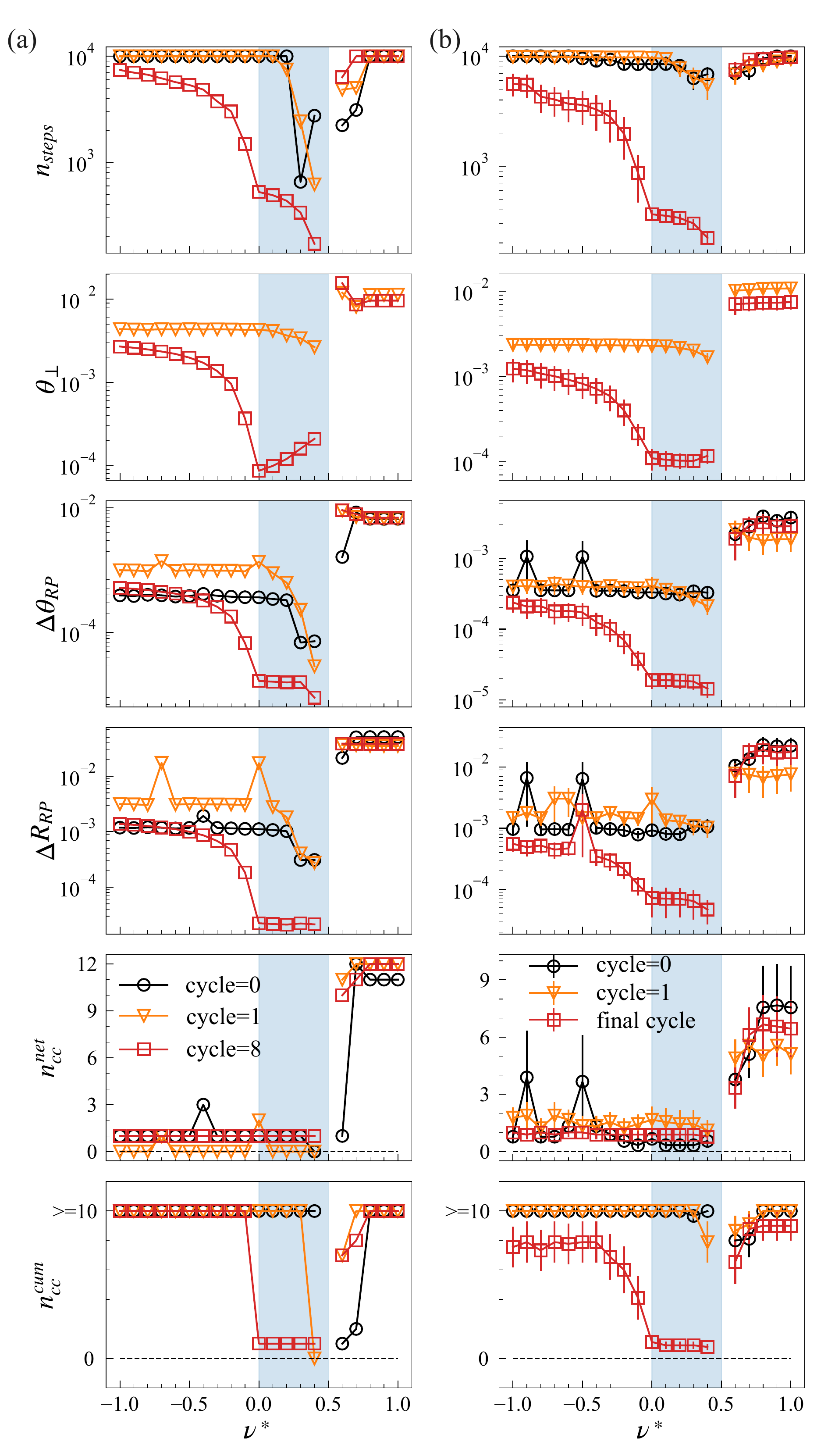}
    \caption{Robustness of MAMs. Column (a) shows a second representative training example that is identical to Fig.~\ref{fig:memory}a except it starts with a different particle configuration. The only qualitative difference in the results is that one contact changes during readout within the training range, corresponding to a Type 1 GD, as expected. 
    Column (b) shows averaged measurements over 10 independent training samples starting from different initial packings. The error bars represent the standard error on the mean.}
    \label{fig:appendix1}
\end{figure}

\section{Ensemble-averaged readout measurements \label{sec:ensemle_average}}
To demonstrate the robustness of marginally absorbing manifolds (MAMs) and \rev{return-point}{physical} memory\rev{ (RPM)}{}, Fig.~\ref{fig:appendix1} shows results, similar to Fig.~\ref{fig:memory}, for an additional representative system trained between Poisson's ratios $\numin=0.0$ and $\numax=0.5$, along with ensemble-averaged data over 10 independent samples. Figure~\ref{fig:appendix1}a replicates the training protocol used in Fig.~\ref{fig:memory}a, but starting from a different initial packing configuration. After 8 cycles, the system converges to a MAM, and the readout for the final state displays qualitatively similar memory behavior as discussed in the main text. The only notable difference is the presence of contact changes between $\numin=0.0$ and $\numax=0.5$, which corresponds to a Type 1 GD as expected.

Figure~\ref{fig:appendix1}(b) shows the same data but averaged over 10 different systems. 
Clear kinks and jumps in the various quantities 
appear at both edges of the training range, highlighting that the systems fall into a MAM and encode the training range. The consistency of these signatures across different realizations underscores the robustness of memory formation in disordered packings under cyclic training. These ensemble results emphasize that memory is not only encoded at the individual-system level but is also statistically reproducible across an ensemble. 

Moreover, outside the training interval, we observe an asymmetry in the system’s ability to adapt: absorbing states obtained through cyclic training are more readily reoptimized toward lower Poisson’s ratios than toward higher ones. Understanding the origin of this asymmetry may be related to the low volume fractions required for systems to reach high values of $\nu$, and will be explored in future work.

\begin{figure}
    \centering
    \includegraphics[width=0.9\linewidth]{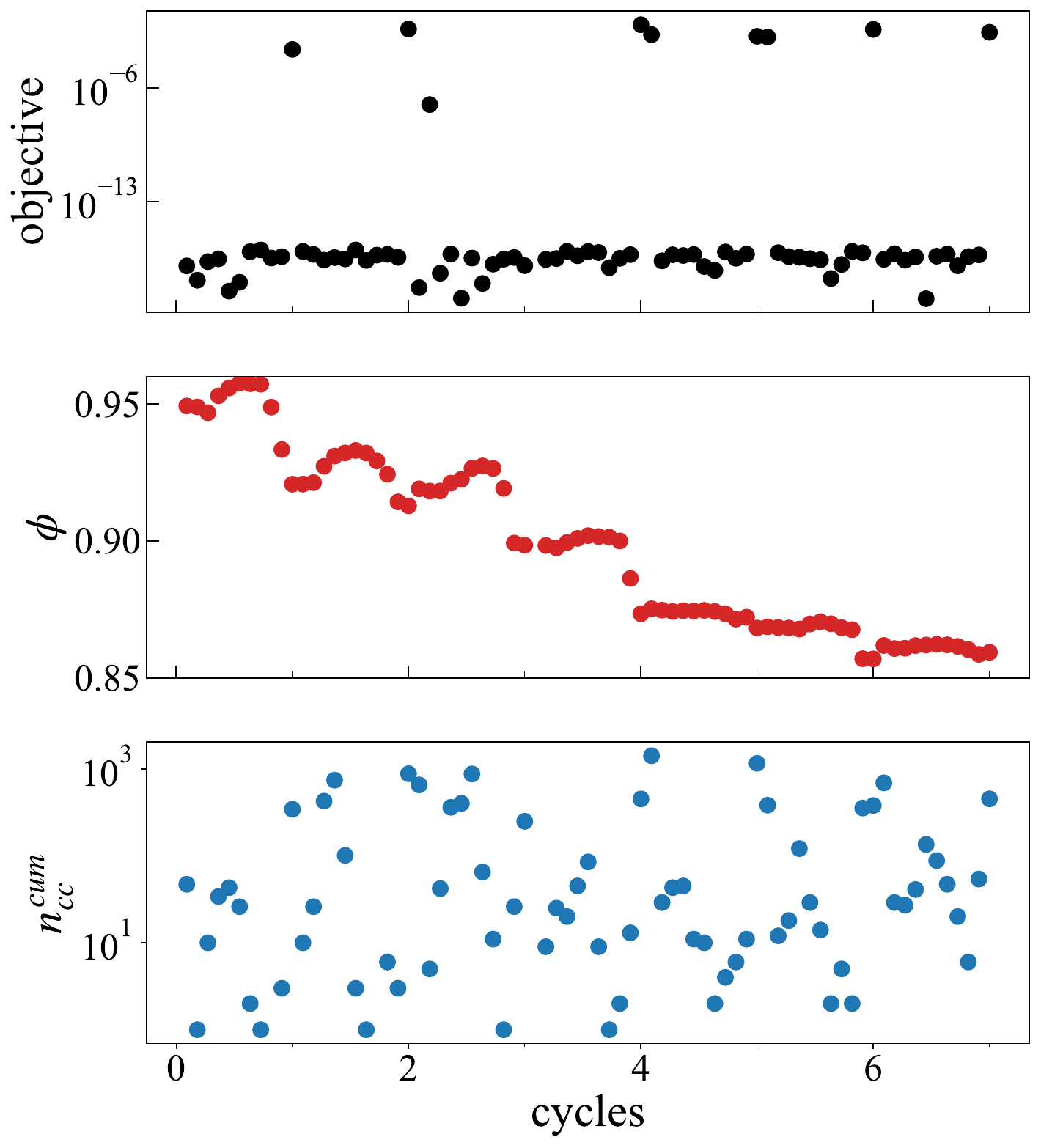}
    \caption{Cyclicly training the Poisson's ratio between $\numax=1.0$ and $\numin=0.5$. From top to bottom, the plot illustrates the evolution of the objective function, measured area fraction $\phi$, and the cumulative contact changes during training $n^{cum}_{\text{cc}}$. After 7 training cycles, the system loses mechanical stability and becomes unjammed, preventing a well-defined calculation of the Poisson's ratio.}
    \label{fig:appendix2}
\end{figure}

\begin{figure}
    \centering
    \includegraphics[width=0.6\linewidth]{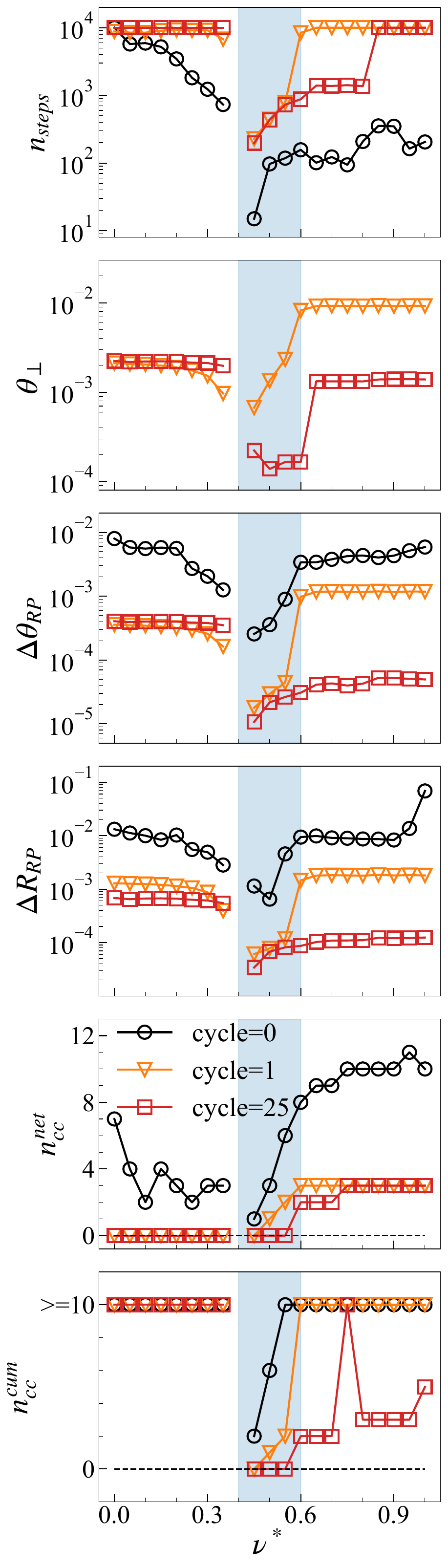}
    \caption{Memory in a representative example trained on a range between $\numax=0.6$ and $\numin=0.4$. The initial Poisson's ratio of the system is $\nu_0=0.605$.}
    \label{fig:appendix3}
\end{figure}

\section{Limits for large Poisson's ratio}
We now discuss the limitations of our model in the regime of large Poisson's ratios. When cyclic training is performed between $\numax=1.0$ and $\numin=0.5$, we observe a pronounced decrease in particle diameters. This reduction leads to a significant drop in area fraction from 0.95 to 0.86 as shown in Fig.~\ref{fig:appendix2}, approaching the jamming threshold $\phi_J=0.848$. In our model, particles interact via a finite-range purely repulsive potential, which results in more frequent structural rearrangements near the jamming point. This dynamic is reflected in the cumulative number of contact changes $n^{cum}_{\text{cc}}$, shown in Fig.~\ref{fig:appendix2}, which exhibits large fluctuations without a clear trend toward convergence. Consequently, cyclic training in this regime becomes increasingly unstable and fails to reliably reach a marginally absorbing manifold.

Despite these challenges at high Poisson's ratios, we demonstrate that training remains feasible for moderately large values. Specifically, as shown in Fig.~\ref{fig:appendix3}, when cyclic training is performed between  $\numax=0.6$ and $\numin=0.4$, the system successfully converges to a MAM after approximately 25 cycles and retains a memory of the training range. While the number of optimization steps required to reach $\nu^*=0.4$ from the initial state obtained via individual training is smaller than that required after cyclic training (black circles), the optimization still involves significant structural rearrangements.

These findings suggest that the model's performance becomes increasingly sensitive to packing constraints at higher $\nu^*$. Further improvements to the optimization procedure—such as incorporating area-fraction constraints directly into the objective function—could enhance stability in this regime. 

\section{Cyclic training on multiple features \label{sec:multiple_features}}
In the main text, we focus on cyclic training targeting a single mechanical feature—specifically, either the Poisson’s ratio or the arbitrarily chosen elastic constant $c_{\text{xxyy}}$. Here, we extend the framework to explore cyclic training with multiple features, which leads to more complex and intriguing characteristics of the resulting absorbing manifold.

We take a randomly generated mechanically stable packing of $N=64$ particles, in which particles are evenly divided into $n_{sp}=16$ species, such that $16$ independent particle diameters serve as tunable parameters. The system is first trained to achieve a target value of the elastic constant $c_{\text{xxyy}}$, beginning at $c^*_{\text{xxyy,min}}=0.12$. Then, training proceeds sequentially by incrementing the target value, using the optimized configuration and parameters from the previous step as the starting point for the next, until reaching $c^*_{\text{xxyy,max}}=0.16$. 
We then reverse the training sequence, returning to $c^*_{\text{xxyy,min}}=0.12$. 

After this full cycle of training  $c_{\text{xxyy}}$, we switch to a second mechanical feature,  $c_{\text{xxxx}}$, and apply a similar cyclic sequence for the target $c^*_\text{xxxx}$: $0.2\rightarrow0.21...\rightarrow0.23\rightarrow0.22...\rightarrow0.2$. We then alternate back and forth, so that a single training cycle involves a complete cycle in each quantity: training in $c_{\text{xxyy}}$ is followed by training in  $c_{\text{xxxx}}$, forming a combined cyclic protocol involving both features. Note that while training one of the target features, the other feature is not constrained and is free to vary.

\begin{figure}
    \centering
    \includegraphics[width=1.\linewidth]{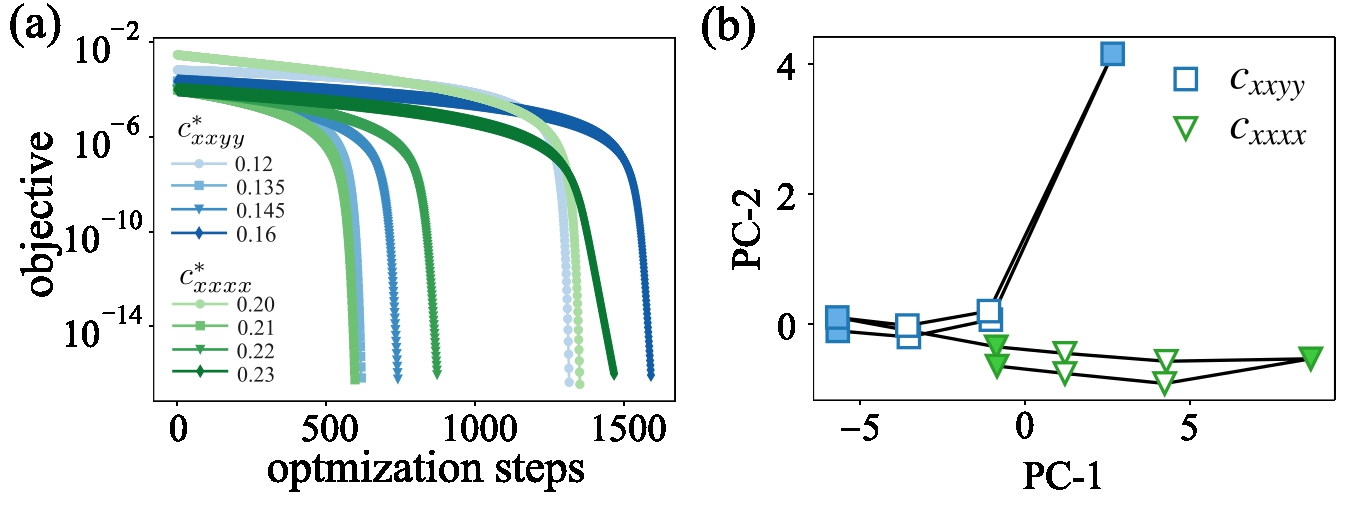}
    \caption{Cyclic training on two elements of elastic modulus tensor. (a) The objective function during a half-cycle of training following 15 combined training cycles. The curves refers to sequential training steps: from $c^*_{\text{xxxx}}=0.2$ to $c^*_{\text{xxyy}}=0.12$, following from $c^*_{\text{xxyy}}=0.12$ to $c^*_{\text{xxyy}}=0.135$ until $c^*_{\text{xxyy}}=0.16$, then from $c^*_{\text{xxyy}}=0.16$ to $c^*_{\text{xxxx}}=0.2$, up to $c^*_{\text{xxxx}}=0.23$. 
    (b) Projection of the parameter vectors onto the first two \rev{principle}{principal} components for the final cycle. The filled symbols refers to the maximum and minimum targets.}
    \label{fig:appendix4}
\end{figure}

After 15 such combined training cycles, the system converges to a MAM, as shown in Fig.~\ref{fig:appendix4}a. To further characterize the structure of this absorbing state, we perform a \rev{principle}{principal} component analysis (PCA) on the parameter evolution during the final cycle. Interestingly, the results, shown in Fig~\ref{fig:appendix4}b, reveal two distinct trajectories in the space of the first two \rev{principle}{principal} components, each corresponding to training on one of the two target features. This separation suggests that memory encoding for each feature follows a distinct pathway in parameter space. Despite this, the switch from training $c_{\text{xxyy}}$ to $c_{\text{xxxx}}$ shows a change along PC-1, which may reflect the coupling or competition between these two features. These preliminary results motivate further quantitative investigations into the nature of memory in systems trained for multiple features.

\bibliography{reference2}

\end{document}